\documentclass[11pt]{article}

\usepackage{amsmath}

\newcommand{\N}{N\raise.7ex\hbox{\underline{$\circ $}}$\;$}

\begin{document}

\begin{center}
{\bf  $4\times 4$ MATRICES   IN DIRAC PARAMETRIZATION:\\ INVERSION
PROBLEM AND DETERMINANT
     \\ [3mm]
     Red'kov V.M., Bogush A.A., Tokarevskaya N.G. \footnote{   E-mail:
redkov@dragon.bas-net.by }\\[3mm]
Institute of Physics, National  Academy of Sciences
of Belarus \\
68 Nezavisimosti  ave.,  Minsk,  BELARUS }

\end{center}

\begin{quotation}

Parametrization of  complex $4 \times 4$ -- matrices $G$   in
terms of Dirac tensor parameters $(A,B,A_{l},B_{l},F_{kl})$ or
equivalent four  complex 4-vectors $(k,m,n,l)$ is investigated. In
the given parametrization, the problem of inverting any $4\times
4$ matrix $G$ is solved. Expression for determinant of any matrix
$G$ is found: $\mbox{det}\; G = F(k,m,n,l)$.

\end{quotation}



\section{
Introduction}

In the context of group theory, the Dirac matrices-based approach
was widely used in physical context:
Macfarlane \cite{1962-Macfarlane}-\cite{1966-Macfarlane},
Hermann \cite{1966-Hermann},
Kilhberg  \cite{1966-Kilhberg}.
 Mack-Todorov \cite{1968-Mack-Todorov},
ten  Kate \cite{1969-Kate},
 Mack -- Salam \cite{1969-Mack-Salam },
 arut - Bohm \cite{1970-Barut-Bohm},
 Mack \cite{1997-Mack}, Barut
-- Bracken  \cite{1975-Bracken}-\cite{1981-Bracken}.
Barut --  Zeni  --   Laufer  \cite{1994-Barut-Zeni-Laufer},
Gsponer  \cite{2002-Gsponer},
Ramakrishna -- Costa \cite{2004-Ramakrishna-Costa}.
However, usually they exploit only general properties of Dirac basis
to parameterize $4\times 4$ matrices. In the present paper we consider three
problems linked to Dirac matrices based approach:

\begin{quotation}

1) Dirac matrix basis  and multiplication in   $GL(4.C)$

2) Inverse matrix  $G^{-1}$

3) Determinant  $\mid G \mid $ in the Dirac parameters

\end{quotation}

\noindent
the problems are rather labarous   technically, but result seem to be important for applications.

\section{ Dirac matrix basis  and multiplication in   $GL(4.C)$}

\hspace{5mm} Any complex matrix  $G \in GL(4.C)$ can be resolved
in terms of 16  Dirac matrices:
\begin{eqnarray}
G = A \; I + iB \; \gamma^{5} + iA_{l}\; \gamma^{l} + B_{l}\;
\gamma^{l} \gamma^{5} + F_{mn} \; \sigma_{mn} \; , \label{A.1}
\end{eqnarray}

\noindent the notation is used
 \begin{eqnarray}
\gamma^{a} \gamma^{b} + \gamma^{b} \gamma^{a} = 2 g^{ab}, \qquad
g^{ab} = \mbox{diag} (+1,-1,-1,-1) \;, \nonumber
\\
\gamma^{5} = -i \gamma^{0}  \gamma^{1} \gamma^{2} \gamma^{3} \; ,
\qquad \sigma^{ab} = {1 \over 4}\; ( \gamma^{a} \gamma^{b} -
\gamma^{b} \gamma^{a}) \; . \label{A.2}
\end{eqnarray}

\noindent  16 coefficients may be taken as independent parameters
in $GL(4.C)$. To establish the composition law for parameters
one should multiply any two matrices of the type  (\ref{A.1}) and
the result obtained is to be decomposed  again in terms of Dirac
matrices:
\begin{eqnarray}
G' = A'\; I + iB '\; \gamma^{5} + iA_{k}'\; \gamma^{k} + B_{k}'\;
\gamma^{k} \gamma^{5} + F'_{cd} \; \sigma^{cd} \; , \nonumber
\\
G = A\; I  + iB \; \gamma^{5} + iA_{l}\; \gamma^{l} + B_{l}\;
\gamma^{l} \gamma^{5} + F_{mn} \; \sigma^{mn} \; , \nonumber
\\
G'' = G'G =  A''\; I + iB ''\; \gamma^{5} + iA_{l}''\; \gamma^{l}
+ B_{l}''\; \gamma^{l} \gamma^{5} + F''_{mn} \; \sigma_{mn} \; .
\nonumber
\\[3mm]
G'' = I \; A'A + iA'B \; \gamma^{5} + i A'A_{k}\; \gamma^{k} +
 A'B_{k} \; \gamma^{k} \gamma^{5} + A'F_{cd} \; \sigma^{cd}
\nonumber
\\
+ iB'A \; \gamma^{5} - B'B \; I + B'A_{k}  \; \gamma^{k}
\gamma^{5}
 -iB'B_{k} \; \gamma^{k} + i B'F_{cd}\; \sigma^{cd}\gamma^{5}
\nonumber
\\
+iA'_{k}A \; \gamma^{k} - A'_{k}B\; \gamma^{k}\gamma^{5} -
 A'_{k}A_{l} \; \gamma^{k}\gamma^{l} +
iA'_{k} B_{l}\; \gamma^{k}\gamma^{l}\gamma^{5} + iA'_{l}F_{cd}\;
\gamma^{l} \sigma^{cd} \nonumber
\\
+ B'_{k} A \; \gamma^{k} \gamma^{5} + iB'_{k}B \; \gamma^{k} -i
B'_{k} A_{l}\; \gamma^{k}\gamma^{l}\gamma^{5} - B'_{k}B_{l} \;
\gamma^{k}\gamma^{l} + B'_{l}F_{cd} \; \gamma^{l}
\sigma^{cd}\gamma^{5} \nonumber
\\
+ F'_{mn} A \; \sigma^{mn}  + i F'_{mn} B \; \sigma^{mn}\gamma^{5}
+
 i F'_{mn} A_{k} \; \sigma^{mn}\gamma^{k}
\nonumber
\\
 +
F'_{mn} B_{k} \; \sigma^{mn} \gamma^{k} \gamma^{5} + F'_{mn}
F_{cd} \; \sigma^{mn} \sigma^{cd} \; . \label{A.5}
\end{eqnarray}

We need some subsidiary relations, they are well known but for
more completeness let us specify some details. The main formula,
base for calculation with Dirac matrices, look as follows
\begin{eqnarray}
\gamma^{a} \gamma^{b} \gamma^{c} = \gamma^{a} g^{bc} - \gamma^{b}
g^{ac} + \gamma^{c} g^{ab} + i\gamma^{5}\;  \epsilon^{abcd}
\;\gamma_{d} \; , \nonumber
\\
\gamma^{5} = -i \gamma^{0}  \gamma^{1} \gamma^{2} \gamma^{3} = {i
\over 24}\; \epsilon_{abcd}\;  \gamma^{a} \gamma^{b} \gamma^{c}
\gamma^{d} \; , \qquad
 \epsilon^{0123}  = +1\;   .
 \label{A.6}
\end{eqnarray}

There are several evident  formulas:
\begin{eqnarray}
\gamma^{a} \gamma^{b} = I \; g^{ab} + 2 \sigma^{ab} \; ;
\nonumber
\end{eqnarray}

\noindent also
\begin{eqnarray}
 \sigma^{ab} \; \gamma^{5} = -{i\over 2} \;
\epsilon^{abkl} \; \sigma_{kl} \; , \qquad
 \gamma^{5}\; \sigma^{ab} =
-{i\over 2} \; \epsilon^{abkl} \; \sigma_{kl} \; ; \label{A.8}
\end{eqnarray}

\noindent  also
\begin{eqnarray}
\gamma^{a} \gamma^{b} \; \gamma^{5} =  ( g^{ab} + 2 \sigma^{ab}
)\; \gamma^{5} = g^{ab} \; \gamma^{5}  - i\;  \epsilon^{abkl} \;
\sigma_{kl} \; ,
\nonumber
\\
\gamma^{5} \; \gamma^{a} \gamma^{b}   =  \gamma^{5}\;  ( g^{ab} +
2 \sigma^{ab} )\;  = g^{ab} \; \gamma^{5}  - i\;  \epsilon^{abkl}
\; \sigma_{kl} \; . \label{A.9b}
\end{eqnarray}

\noindent From identity
\begin{eqnarray}
\gamma^{l} \sigma^{cd} = {1 \over 4 } \gamma^{l} \; (\gamma^{c}
\gamma^{d} - \gamma^{d} \gamma^{c}  ) = \nonumber
\\
= {1 \over 4} \; \left [ \; \gamma^{l} g^{cd} - \gamma^{c} g^{ld}
+  \gamma^{d} g^{lc} + i \gamma^{5}
 \epsilon^{lcds}\gamma_{s}-
\gamma^{l} g^{dc} + \gamma^{d} g^{lc} -  \gamma^{c} g^{ld} - i
\gamma^{5}
 \epsilon^{ldcs}\gamma_{s}\;
\right ] , \nonumber
\end{eqnarray}

\noindent it follows
\begin{eqnarray}
\gamma^{l} \sigma^{cd} = {1 \over 2} \; \left [ \;
  g^{lc}  \gamma^{d} - g^{ld}\gamma^{c}  + i \gamma^{5}
 \epsilon^{lcds}\gamma_{s} \; \right ] ,
\label{A.10a}
\end{eqnarray}

\noindent in the same manner
\begin{eqnarray}
\sigma^{mn} \gamma^{k}  = {1 \over 2} \; \left [ \; \gamma^{m}
g^{nk}   - \gamma^{n}  g^{mk}   + i \gamma^{5}
 \epsilon^{mnks}\gamma_{s} \; \right ] .
\label{A.10b}
\end{eqnarray}

\noindent There are two similar  formulas with involved
$\gamma^{5}$:
\begin{eqnarray}
\gamma^{l}  \sigma^{cd}  \gamma^{5} = {1 \over 2} \; \left [ \;
  g^{lc}  \gamma^{d}  \gamma^{5} - g^{ld}\gamma^{c}  \gamma^{5}  -i
 \epsilon^{lcds}\gamma_{s} \; \right ] ,
\nonumber
\\
 \sigma^{mn} \gamma^{k} \gamma^{5}   =
{1 \over 2} \; \left [ \; \gamma^{m}  \gamma^{5} g^{nk}   -
\gamma^{n} \gamma^{5} g^{mk}   - i
 \epsilon^{mnks}\gamma_{s} \; \right ] .
\label{A.11b}
\end{eqnarray}

\noindent Finally, we need one  other  combination
\begin{eqnarray}
\sigma^{mn} \sigma^{cd} = {1 \over 4}\;  \; \gamma^{m} \;
(\gamma^{n}  \sigma^{cd} ) - {1 \over 4}\; \gamma^{n} \; (
\gamma^{m} \; \sigma^{cd}) \; \nonumber
\\
= {1 \over 8}\; \left [   \; \gamma^{m} \;  (\; g^{nc} \gamma^{d}
- g^{nd} \gamma^{c} +i \gamma^{5} \epsilon^{ncds} \gamma_{s} \;) -
\gamma^{n} \;  (\; g^{mc} \gamma^{d}  - g^{md} \gamma^{c} +i
\gamma^{5} \epsilon^{mcds} \gamma_{s} \;) \; \right ] \nonumber
\\
=
 {1 \over 8}\; \left [ \;
 ( g^{nc} \gamma^{m}\gamma^{d} - g^{nd} \gamma^{m}\gamma^{c})-
( g^{mc} \gamma^{n}\gamma^{d} - g^{md} \gamma^{n}\gamma^{c}) - (i
\epsilon ^{ncd}_{\;\;\;\;\;\;\;s}\; \gamma^{m}\gamma^{s}
\gamma^{5} -i \epsilon ^{mcd}_{\;\;\;\;\;\;\;s}\;
\gamma^{n}\gamma^{s} \gamma^{5} )\; \right ]\; . \nonumber
\end{eqnarray}

\noindent which  gives
\begin{eqnarray}
\sigma^{mn} \sigma^{cd} =
 {1 \over 8}\;  \{ \;  [ \;
   g^{nc} (g^{md} + 2 \sigma^{md})  - g^{nd} (g^{mc} + 2 \sigma^{mc}) \; ]
   \nonumber
   \\
   -
 [\; ( g^{mc}  ( g^{nd} + 2 \sigma^{nd} ) - g^{md} ( g^{nc} +
2 \sigma^{nc})\;  ] \nonumber
\\
 -   [ \;i\epsilon ^{ncd}_{\;\;\;\;\;\;s}\; (g^{ms}
\gamma^{5} - i \epsilon^{mskl} \sigma_{kl})  -i \epsilon
^{mcd}_{\;\;\;\;\;\;s}\; (g^{ns} \gamma^{5} -
 i \epsilon^{nskl} \sigma_{kl}) \;  ] \;  \}
\nonumber
\\
= {1 \over 8}\;  \{ \;  [ \;
   (g^{nc} g^{md} - g^{nd} g^{mc} ) - (g^{mc} g^{nd} - g^{md} g^{nc} )\;  ]
\nonumber
\\
  + 2 [ \;( g^{nc} \sigma^{md}   -  g^{nd} \sigma^{mc})  -
(  g^{mc}  \sigma^{nd}  -  g^{md} \sigma^{nc} ) \;  ] +2i\;
\epsilon^{mncd} \; \gamma^{5} \nonumber
\\
 -   ( \; \epsilon^{ncd}_{\;\;\;\;\;\;s} \;
\epsilon^{mkls} \; \sigma_{kl}  - \epsilon^{mcd}_{\;\;\;\;\;\;s}
\; \epsilon^{nkls} \; \sigma_{kl}\;  ) \;   \}   . \nonumber
\end{eqnarray}

\noindent Further, using  the known identity
\begin{eqnarray}
\epsilon^{ncd}_{\;\;\;\;\;\;s} \; \epsilon^{mkls}  \; \sigma_{kl}
= -   \left | \begin{array}{ccc}
g^{mn} & g^{mc} & g^{md} \\
g^{kn} & g^{kc} & g^{kd} \\
g^{ln} & g^{lc} & g^{ld}      \end{array} \right | \; \sigma_{kl}
\;, \nonumber
\end{eqnarray}

\noindent after simple calculation we get
\begin{eqnarray}
\sigma^{mn} \sigma^{cd} = {1 \over 8}\;  [ \;
   (g^{nc} g^{md} - g^{nd} g^{mc} ) - (g^{mc} g^{nd} - g^{md} g^{nc} )\;  ]
\nonumber
\\
 +  {i \over 4} \; \epsilon^{mncd} \; \gamma^{5} + {1 \over 2}\;
   [ \;
(g^{nc} \sigma^{md} - g^{nd}\sigma^{mc}) - (g^{mc} \sigma^{nd} -
g^{md}\sigma^{nc}) \;  ]\;. \label{A.12}
\end{eqnarray}

From   (\ref{A.5}) we arrive at
\begin{eqnarray}
G'' = G'G =  A''\; I + iB ''\; \gamma^{5} + iA_{l}''\; \gamma^{l}
+ B_{l}''\; \gamma^{l} \gamma^{5} + F''_{mn} \; \sigma_{mn}
\nonumber
\\
= A'A\; I  + i A'B \; \gamma^{5} + i A' A_{l} \; \gamma^{l}  +
A'B_{l} \; \gamma^{l}\gamma^{5} + A' F_{kl} \; \sigma^{kl}
\nonumber
\\
+ iB'A\; \gamma^{5} - B'B \; I + B'A_{l}\; \gamma^{l}\gamma^{5} -
i B'B_{l}\; \gamma^{l} + i B'F_{cd} \; (-i/2) \; \epsilon^{cdkl}
\; \sigma_{kl}
\nonumber
\\
+iA'_{l}A \; \gamma^{l} - A'_{l}B \; \gamma^{l}\gamma^{5}  -
A'_{l}A_{k} (g^{lk} +2\sigma^{lk}) +iA'_{l}B_{k}\; (g^{lk}
\gamma^{5} - i \;\epsilon^{lkmn}\sigma_{mn}) \nonumber
\nonumber
\end{eqnarray}
\begin{eqnarray}
+ iA'_{l}F_{cd}  \; {1\over 2}\; [ \; (g^{lc}\gamma^{d} -
g^{ld}\gamma^{c})  + i \; \gamma^{5} \epsilon^{lcds}\gamma_{s} \;
] + B'_{l}A \; \gamma^{l} \gamma^{5}  + iB'_{l}B\; \gamma^{l}
\nonumber
\\
-iB'_{l}A_{k} \; (g^{lk} \gamma^{5} - i\; \epsilon^{lkmn}\;
\sigma_{mn} ) -
 B'_{l}B_{k} \; (g^{lk} + 2 \sigma^{lk})
\nonumber
\\
+ B'_{l} F_{cd} \; {1\over 2}\;  [\; ( g^{lc} \gamma^{d}\gamma^{5}
- g^{ld} \gamma^{c}\gamma^{5} ) - i \; \epsilon^{lcds}\;
\gamma_{s} \;  ] + F'_{mn} A\; \sigma^{mn} \nonumber
\\
+ iF'_{mn} B \; (-i/2)\; \epsilon^{mnkl}\; \sigma_{kl} + iF'_{mn}
A_{k} \; {1\over 2}\;  [ \; (\gamma^{m} g^{nk} - \gamma^{n}
g^{mk})  + i \gamma^{5} \; \epsilon^{mnks} \; \gamma_{s} \; ]
\nonumber
\\
+ F'_{mn}B_{k} \; {1 \over 2}\;  [\; (\gamma^{m} \gamma^{5} g^{nk}
- \gamma^{n} \gamma^{5} g^{mk}) - i\; \epsilon^{mnks}\; \gamma_{s}
\;  ] \nonumber
\\
+ F'_{mn}F_{cd}\;  \{ {1 \over 8}\;  [ \;
   (g^{nc} g^{md} - g^{nd} g^{mc} ) - (g^{mc} g^{nd} - g^{md} g^{nc} )\;  ]
\nonumber
\\
  +  {i \over 4} \; \epsilon^{mncd} \; \gamma^{5} + {1 \over
2}\;
  [ \;
(g^{nc} \sigma^{md} - g^{nd}\sigma^{mc}) - (g^{mc} \sigma^{nd} -
g^{md}\sigma^{nc}) \;  ]  \}\;. \label{A.13}
\end{eqnarray}

\noindent In the first place, expression for two scalars are
produced:
\begin{eqnarray}
A''= A'\; A -B'\; B -A'_{l} \; A^{l} -B'_{l} \; B^{l} - {1\over 2
} \;F'_{kl} \;F^{kl} \; ,
\nonumber
\\
B''= A'\; B + B' \;A +  A'_{l} \; B^{l} - B'_{l} \; A^{l} +
{1\over 4 }\; F'_{mn} \;F_{cd}\;
 \epsilon^{mncd}\; .
\label{A.14b}
\end{eqnarray}

\noindent Now, from
\begin{eqnarray}
i A''_{l} \; \gamma^{l} = i A'A_{l} \; \gamma^{l} -iB'B_{l} \;
\gamma^{l} + i A'_{l}A \; \gamma^{l} + i A'_{l} F_{cd} \; {1\over
2}\;(g^{lc} \gamma^{d} - g^{ld} \gamma^{c} ) + i B'_{l}B \;
\gamma^{l} \nonumber
\\
- i B'_{l}F_{cd}\; {1\over 2}\; \epsilon^{lcds}\;\gamma_{s} + i
F'_{mn} A_{k} \; {1\over 2} \;(\gamma^{m} g^{nk} - \gamma^{n}
g^{mk}) - {i\over 2}\; F'_{mn}B_{k} \; \epsilon^{mnks}
\;\gamma_{s} \;  , \nonumber
\end{eqnarray}

\noindent and
\begin{eqnarray}
\gamma^{l} \gamma^{5} \; B''_{l} = A'B_{l} \; \gamma^{l}
\gamma^{5}   +   B' A_{l} \; \gamma^{l} \gamma^{5} - A'_{l}B\;
\gamma^{l}\gamma^{5} + A'_{l} F_{cd} \; {1 \over 2} \;
\epsilon^{lcds} \; \gamma_{s}\gamma^{5} + B'_{l} A \; \gamma^{l}
\gamma^{5} \nonumber
\\
+ B'_{l} F_{cd} \; {1 \over 2}\; ( g^{lc}g^{d}\gamma^{5} -
g^{ld}g^{c}\gamma^{5} ) + {1\over 2} \; F'_{mn} A_{k}\;
\epsilon^{mnks} \; \gamma_{s} \gamma^{5}  + {1\over 2}\; F'_{mn}
B_{k} \; ( \gamma^{m} \gamma^{5} g^{nk} - \gamma^{n} \gamma^{5}
g^{mk} ) \; . \nonumber
\end{eqnarray}

\noindent it follow expressions for   $A''_{l}$  and  $B''_{l}$:
\begin{eqnarray}
A''_{l} = A' \;  A_{l} - B' \; B_{l} + A'_{l} \; A + B'_{l} \; B +
A'^{k} \; F_{kl} \nonumber
\\
+ F'_{lk} \; A^{k} + {1\over 2} \; B'_{k} \; F_{mn} \;
\epsilon_{l}^{\;\;\;kmn} + {1\over 2}\; F'_{mn} \; B_{k} \;
\epsilon_{l}^{\;\;\;mnk} \; ; \label{A.15a}
\\[2mm]
 B''_{l} = A' \; B_{l} + B' \; A_{l} - A'_{l} \; B + B'_{l} \;  A + B'^{k} \; F_{kl}
 \nonumber
 \\
 +  F'_{lk} \;  B^{k} + {1\over 2} \; A'_{k} \;  F_{mn}\; \epsilon^{kmn}_{\;\;\;\;\;\;\;l} +
 {1\over 2} \; F'_{mn} \;  A_{k} \; \epsilon^{mnk}_{\;\;\;\;\;\;\;l} \; .
\label{A.15b}
\end{eqnarray}

 Finally, because
\begin{eqnarray}
 \sigma^{mn} \;F_{mn} = A'F_{mn} \sigma^{mn} +
 {1\over 2}\;  B'F_{cd}\;  \epsilon^{cdmn} \; \sigma_{mn}  - 2A'_{m}A_{n} \; \sigma^{mn} +
  A'_{l}B_{k} \; \epsilon^{lkmn}\; \sigma^{mn}
\nonumber
\\
  -
  B'_{l} A_{k} \;  \epsilon^{lkmn}\; \sigma_{mn} -
  2B'_{m}B_{n} \; \sigma^{mn} + F'_{mn} A \; \sigma^{mn}
      +
   {1\over 2}\; F'_{kl} B\; \epsilon^{klmn}\; \sigma_{mn}
\nonumber
\\
   +
   {1\over 2}\;
   F'_{mn}F_{cd}\;
    [ \; (g^{nc}\sigma^{md} - g^{nd}\sigma^{mc})-
(g^{mc}\sigma^{nd} - g^{md}\sigma^{nc})\;  ] \; , \nonumber
\end{eqnarray}

\noindent   the  tensor quantity  $F''_{mn}$ is
\begin{eqnarray}
F''_{mn} = A' \; F_{mn} + F'_{mn} \;  A  - (A'_{m} \; A_{n}
-A'_{n} \; A_{m}) - (B'_{m} \; B_{n} -B'_{n} \; B_{m}) \nonumber
\\
+ A'_{l} \; B_{k} \; \epsilon^{lkmn} - B'_{l} \;  A_{k} \;
\epsilon^{lkmn}\; +{1\over 2} \;  B' \; F_{kl} \;
\epsilon^{kl}_{\;\;\;\;\;mn} + {1\over 2} \; F'_{kl}\; B\;
\epsilon^{kl}_{\;\;\;\;\;mn} \nonumber
\\
+ (F'_{mk} \; F^{k}_{\;\;n} -  F'_{nk} \;  F^{k}_{\;\;m}  )\; .
\label{A.16}
\end{eqnarray}

Thus, multiplication law for the group $GL(4.C)$,  and all its
sub-groups  are described by  one the same formula:
\begin{eqnarray}
G'G =  A''\; I + iB ''\; \gamma^{5} + iA_{l}''\; \gamma^{l} +
B_{l}''\; \gamma^{l} \gamma^{5} + F''_{mn} \; \sigma_{mn} \; ,
\nonumber
\\[3mm]
A''= A'\; A -B'\; B -A'_{l} \; A^{l} -B'_{l} \; B^{l} - {1\over 2
} \;F'_{kl} \;F^{kl} \; , \nonumber
\\
B''= A'\; B + B'\; A +  A'_{l} \; B^{l} - B'_{l} \; A^{l} +
{1\over 4 } \;F'_{mn} \;F_{cd}\;
 \epsilon^{mncd}\; ,
\nonumber
\\
A''_{l} = A' \;A_{l} - B'\;B_{l} + A'_{l\;}A + B'_{l}\;B  +
A'^{k}F_{kl} \nonumber
\\
+ F'_{lk}A^{k} + {1\over 2} \; B'_{k} \;F_{mn} \;
\epsilon_{l}^{\;\;\;kmn} + {1\over 2}\; F'_{mn}\;B_{k} \;
\epsilon_{l}^{\;\;\;mnk} \; , \nonumber
\\
 B''_{l} = A'\;B_{l} + B'\;A_{l} - A'_{l}\;B + B'_{l} \;A + B'^{k}\;F_{kl}
\nonumber
\\
 +  F'_{lk} \;B^{k} + {1\over 2} \; A'_{k} \;F_{mn}\; \epsilon^{kmn}_{\;\;\;\;\;\;\;l} +
 {1\over 2} \; F'_{mn} \;A_{k} \; \epsilon^{mnk}_{\;\;\;\;\;\;\;l} \; ,
\nonumber
\\
F''_{mn} = A'\;F_{mn} + F'_{mn} \;A  - (A'_{m} \; A_{n} -A'_{n} \;
A_{m}) - (B'_{m}\;B_{n} -B'_{n}\;B_{m}) \nonumber
\\
+ A'_{l}\;B_{k} \; \epsilon^{lkmn} - B'_{l}\; A_{k} \;
\epsilon^{lkmn}\; +{1\over 2} \;  B'\;F_{kl} \;
\epsilon^{kl}_{\;\;\;\;\;mn} + {1\over 2} \; F'_{kl}\;B\;
\epsilon^{kl}_{\;\;\;\;\;mn} \nonumber
\\
+ (F'_{mk} \;F^{k}_{\;\;n} -  F'_{nk} \;F^{k}_{\;\;m}  )\; .
\label{A.17}
\end{eqnarray}

\section{ Inverse matrix  $G^{-1}$ }

Let a  matrix   $G$ is given by
\begin{eqnarray}
G = \left | \begin{array}{cccc}
+(k_{0} + k_{3})    &  +(k_{1} - ik_{2})  & \hspace{3mm}  +(n_{0} - n_{3})    &  -(n_{1} - in_{2}) \\
+(k_{1} + ik_{2} )  & + (k_{0} - k_{3} )  &  \hspace{3mm}-(n_{1} + in_{2})   &  + (n_{0} + n_{3} ) \\[3mm]
-(l_{0} + l_{3})   & -(l_{1} - il_{2})  & \hspace{3mm} +(m_{0} - m_{3})    &  -(m_{1} - im_{2}) \\
-(l_{1} + il_{2})   & -(l_{0} - l_{3})  &  \hspace{3mm} -(m_{1} +
im_{2})   & +(m_{0} + m_{3})
\end{array} \right | \; ,
\label{B.1a}
\end{eqnarray}

\noindent For inverse matrix we have general expression
\begin{eqnarray}
G^{-1} =  \mid G \mid^{-1} \; \left | \begin{array}{cccc}
A_{11} & A_{21} &  A_{31} & A_{41} \\
A_{12} & A_{22} &  A_{32} & A_{42} \\
A_{13} & A_{23} &  A_{33} & A_{43} \\
A_{14} & A_{24} &  A_{34} & A_{44}
\end{array} \right | \; .
\label{B.1b}
\end{eqnarray}

Let us find   $(k_{0})^{-1} $ and   $(k_{3})^{-1}$:
\begin{eqnarray}
{A_{11} + A_{22} \over 2 \; \mbox{det}\; G } = (k_{0})^{-1} \; ,
\qquad {A_{11} - A_{22} \over 2 \; \mbox{det}\; G } = (k_{3})^{-1}
\; . \label{B.2}
\end{eqnarray}

\noindent Cofactor  $A_{11}$ is
\begin{eqnarray}
A_{11} = \left | \begin{array}{rrr}
 k_{0} - k_{3}   & -n_{1} - in_{2}   &   n_{0} + n_{3}  \\
 -l_{1} + il_{2}  &  m_{0} - m_{3}    &  -m_{1} + im_{2} \\
 -l_{0} + l_{3}  & -m_{1} - im_{2}   &   m_{0} + m_{3}
\end{array} \right | =
\nonumber
\\
= (k_{0} -k_{3}) \; (mm) +
 (n_{0} +n_{3}) (l_{1} -il_{2})(m_{1} + im_{2})
\nonumber
\\
- (l_{0}-l_{3})( n_{1} + i n_{2})( m_{1} -im_{2}) \nonumber
\\
- (m_{0} + m_{3})( l_{1} -il_{2})(n_{1} + in_{2}) \nonumber
\\
+ (l_{0}-l_{3})(m_{0}-m_{3})(n_{0} +n_{3}) \; , \label{B.3a}
\end{eqnarray}

\noindent Cofactor  $A_{22}$  is
\begin{eqnarray}
A_{22} = \left | \begin{array}{rrr}
k_{0} + k_{3}      &  n_{0} - n_{3}    &  -n_{1} + in_{2} \\
-l_{0} - l_{3}     &  m_{0} - m_{3}    &  -m_{1} + im_{2} \\
-l_{1} - il_{2}   & -m_{1} - im_{2}   &   m_{0} + m_{3}
\end{array} \right | =
\nonumber
\\
= (k_{0} +k_{3}) \; (mm) + (n_{0}  -n_{3}) (l_{1} +il_{2})(m_{1} -
im_{2}) \nonumber
\\
- (l_{0}+l_{3})( n_{1} - i n_{2})( m_{1} +im_{2}) \nonumber
\\
- (m_{0} - m_{3})( l_{1} +il_{2})(n_{1} - in_{2}) \nonumber
\\
+ (l_{0}+l_{3})(m_{0}+m_{3})(n_{0} -n_{3}) \; . \label{B.3b}
\end{eqnarray}

\noindent With the use of identities
\begin{eqnarray}
{1 \over 2} \; [  (n_{0} +n_{3}) (l_{1} -il_{2})(m_{1} + im_{2}) +
(n_{0}  -n_{3}) (l_{1} +il_{2})(m_{1} - im_{2})  ] \nonumber
\\
= n_{0} l_{1} m_{1} + n_{0} l_{2} m_{2} + in_{3} l_{1} m_{2} - i
n_{3} l_{2} m_{1} \; , \nonumber
\\
{1 \over 2} \; [  (n_{0} +n_{3}) (l_{1} -il_{2})(m_{1} + im_{2}) -
(n_{0}  -n_{3}) (l_{1} +il_{2})(m_{1} - im_{2})  ] \nonumber
\\
= i n_{0} l_{1} m_{2} -i n_{0} l_{2} m_{1} + n_{3} l_{1} m_{1} +
n_{3} l_{2} m_{2} \; ,
\nonumber
\end{eqnarray}
\begin{eqnarray}
- {1 \over 2} \; [\; (l_{0}-l_{3})( n_{1} + i n_{2})( m_{1}
-im_{2})  + (l_{0}+l_{3})( n_{1} - i n_{2})( m_{1} +im_{2})\; ]
\nonumber
\\
= - l_{0} n_{1}m_{1} - l_{0} n_{2} m_{2} - i l_{3} n_{1} m_{2} + i
l_{3} n_{2} m_{1} \;, \nonumber
\\
- {1 \over 2} \; [\; (l_{0}-l_{3})( n_{1} + i n_{2})( m_{1}
-im_{2})  - (l_{0}+l_{3})( n_{1} - i n_{2})( m_{1} +im_{2})\; ]
\nonumber
\\
= il_{0} n_{1} m_{2} -i l_{0} n_{2} m_{1} + l_{3} n_{1} m_{1} +
l_{3}n_{2}m_{2} \; .
\nonumber
\end{eqnarray}
\begin{eqnarray}
-{1 \over 2}\; [\;  (m_{0} + m_{3})( l_{1} -il_{2})(n_{1} +
in_{2}) +
 (m_{0} - m_{3})( l_{1} +il_{2})(n_{1} - in_{2}) \; ]
\nonumber
\\
= - m_{0} l_{1} n_{1} - m_{0} l_{2} n_{2} - i m_{3} l_{1} n_{2} +
i m_{3} l_{2} n_{1} \; , \nonumber
\\
-{1 \over 2}\; [\;  (m_{0} + m_{3})( l_{1} -il_{2})(n_{1} +
in_{2}) -
 (m_{0} - m_{3})( l_{1} +il_{2})(n_{1} - in_{2}) \; ]
\nonumber
\\
= -i m_{0} l_{1} n_{2} +i m_{0} l_{2} n_{1} -m_{3} l_{1} n_{1} -
m_{3} l_{2} n_{2} \; ,
\nonumber
\end{eqnarray}
\begin{eqnarray}
{1 \over 2} \; [\;  (l_{0}-l_{3})(m_{0}-m_{3})(n_{0} +n_{3}) +
(l_{0}+l_{3})(m_{0}+m_{3})(n_{0} -n_{3})\; ] \nonumber
\\
= l_{0} m_{0} n_{0} - l_{0} m_{3} n_{3}  - l_{3} m_{0} n_{3} +
l_{3} m_{3} n_{0} \; , \nonumber
\\
{1 \over 2} \; [\;  (l_{0}-l_{3})(m_{0}-m_{3})(n_{0} +n_{3}) -
(l_{0}+l_{3})(m_{0}+m_{3})(n_{0} -n_{3})\; ] \nonumber
\\
= l_{0} m_{0} n_{3} - l_{0} m_{3} n_{0} - l_{3} m_{0} n_{0} +
l_{3} m_{3} n_{3} \ ;,
\nonumber
\end{eqnarray}

\noindent  we find $(k_{0})^{-1}$ and $(k_{3})^{-1}$:
\begin{eqnarray}
(k_{0})^{-1}= + k_{0} \; (mm) \nonumber
\\
+ n_{0} l_{1} m_{1} + n_{0} l_{2} m_{2} + in_{3} l_{1} m_{2} - i
n_{3} l_{2} m_{1} \nonumber
\\
- l_{0} n_{1}m_{1} - l_{0} n_{2} m_{2} - i l_{3} n_{1} m_{2} + i
l_{3} n_{2} m_{1} \nonumber
\\
- m_{0} l_{1} n_{1} - m_{0} l_{2} n_{2} - i m_{3} l_{1} n_{2} + i
m_{3} l_{2} n_{1} \nonumber
\\
+ l_{0} m_{0} n_{0} - l_{0} m_{3} n_{3}  - l_{3} m_{0} n_{3} +
l_{3} m_{3} n_{0}\; ,
\nonumber
\end{eqnarray}
\begin{eqnarray}
(k_{3})^{-1}= - k_{3} \; (mm) \nonumber
\\
+ i n_{0} l_{1} m_{2} -i n_{0} l_{2} m_{1} + n_{3} l_{1} m_{1}
n_{3} l_{2} m_{2}  + \nonumber
\\
+i l_{0} n_{1} m_{2} -i l_{0} n_{2} m_{1} + l_{3} n_{1} m_{1} +
l_{3}n_{2}m_{2} \nonumber
\\
-i m_{0} l_{1} n_{2} +i m_{0} l_{2} n_{1} -m_{3} l_{1} n_{1} -
m_{3} l_{2} n_{2} \nonumber
\\
+ l_{0} m_{0} n_{3} - l_{0} m_{3} n_{0} - l_{3} m_{0} n_{0} +
l_{3} m_{3} n_{3} \; .
\nonumber
\end{eqnarray}

\noindent From this, after identical transformations, we arrive at
 (for brevity the factor  $ \mid G \mid ^{-1}$  is omitted)
\begin{eqnarray}
(k_{0})^{-1} = k_{0} \; (mm) + m_{0} \; (ln) + l_{0} \; (nm)
-n_{0} (lm)  \; +
  i \;   {\bf l} \;({\bf m} \times {\bf n} ) \; ,
  \label{B.6a}
  \end{eqnarray}

\noindent where
\begin{eqnarray}
i \;   {\bf l} \;({\bf m} \times {\bf n} ) = i \; [\; l_{1}
(m_{2}n_{3} -m_{3}n_{2}) + l_{2} (m_{3}n_{1} -m_{1}n_{3})+ l_{3}
(m_{1}n_{2} -m_{2}n_{1}) \; ]\;  ; \nonumber
\end{eqnarray}

\noindent and
\begin{eqnarray}
(k_{3})^{-1} = -k_{3} \; (mm) - m_{3} \; (ln) - l_{3} \; (nm) +
 n_{3 } (lm) \;
\nonumber
\\
 + 2 \; [ \; {\bf l} \times ({\bf n} \times {\bf m}) \; ]_{3} \;+
  i\; [\; m_{0} ( {\bf n} \times {\bf l} )_{3}  +
 l_{0} ( {\bf n} \times {\bf m} )_{3} +
n_{0} ( {\bf l} \times {\bf m} )_{3}  \; ] \;, \label{B.6b}
\end{eqnarray}

\noindent where
\begin{eqnarray}
2 \; [ \; {\bf l} \times ({\bf n} \times {\bf m}) \; ]_{3} = 2\;
[\; l_{1}\; (n_{3}  m_{1} -n_{1} m_{3} ) - l_{2} (n_{2}m_{3} -
n_{3} m_{2}) \;]\;  , \nonumber
\\
i\; [\; m_{0} ( {\bf n} \times {\bf l} )_{3}  +
 l_{0} ( {\bf n} \times {\bf m} )_{3} +
n_{0} ( {\bf l} \times {\bf m} )_{3}  \; ] \nonumber
\\
= i \; [\; m_{0} (  n_{1} l_{2} -  n_{2} l_{1}) +
 l_{0} (  n_{1}  m_{2} -  n_{2}  m_{1} ) +
n_{0} ( l_{1} m_{2} - l_{2} m_{1}  )  \; ] \; . \nonumber
\end{eqnarray}

\vspace{5mm} Now, let us find  $(k_{1})^{-1}$  and $(k_{1})^{-1}$:
\begin{eqnarray}
k_{1} = {1 \over 2 \mid G \mid } \; (A_{12} + A_{21}) \; , \qquad
i k_{2} = {1 \over 2 \mid G \mid } \; (A_{12} - A_{21}) \; .
\label{B.7}
\end{eqnarray}

\noindent Cofactor  $A_{12}$ is
\begin{eqnarray}
A_{12} = (-1) \left | \begin{array}{rrr}
k_{1} + ik_{2}      & -n_{1} - in_{2}   &   n_{0} + n_{3}  \\
-l_{0} - l_{3}     &  m_{0} - m_{3}    &  -m_{1} + im_{2} \\
-l_{1} - il_{2}   & -m_{1} - im_{2}   &   m_{0} + m_{3}
\end{array} \right |
\nonumber
\\
=(-1) \{ \; (k_{1} +ik_{2}) \; (mm)  +
(l_{0}+l_{3})(n_{0}+n_{3})(m_{1}+ i m_{2}) \nonumber
\\
+(m_{0} - m_{3}) ( n_{0} +n_{3}) (l_{1} + il_{2}) \nonumber
\\
- (m_{0} +m_{3})( l_{0} +l_{3}) (n_{1} + i n_{2}) \nonumber
\\
- (l_{1} + il_{2})(n_{1} + i n_{2})( m_{1} -i m_{2}) \; \} \; .
\label{B.8a}
\end{eqnarray}

\noindent  Cofactor  $A_{21}$ is
\begin{eqnarray}
A_{21} = (-1) \left | \begin{array}{rrr}
 k_{1} - ik_{2}  &  n_{0} - n_{3}    &  -n_{1} + in_{2} \\
 -l_{1} + il_{2}  &  m_{0} - m_{3}    &  -m_{1} + im_{2} \\
 -l_{0} + l_{3}  & -m_{1} - im_{2}   &   m_{0} + m_{3}
\end{array} \right |
\nonumber
\\
=(-1) \{ \; (k_{1} -ik_{2}) \; (mm) +
(l_{0}-l_{3})(n_{0}-n_{3})(m_{1}- i m_{2}) \nonumber
\\
+(m_{0} - m_{3}) ( n_{0} -n_{3}) (l_{1} - il_{2}) \nonumber
\\
- (m_{0} -m_{3})( l_{0} -l_{3}) (n_{1} - i n_{2}) \nonumber
\\
- (l_{1} - il_{2})(n_{1} - i n_{2})( m_{1} +i m_{2}) \; \} \; .
\label{B.8b}
\end{eqnarray}

\noindent With the use of identities:
\begin{eqnarray}
{1 \over 2} \; [\; (l_{0}+l_{3})(n_{0}+n_{3})(m_{1}+ i m_{2}) +
(l_{0}-l_{3})(n_{0}-n_{3})(m_{1}- i m_{2}) \; ] \nonumber
\\
= l_{0} n_{0}m_{1} + l_{3} n_{3} m_{1} + i l_{0} n_{3} m_{2} + i
l_{3} n_{0} m_{2} \; , \nonumber
\\
{1 \over 2} \; [\; (l_{0}+l_{3})(n_{0}+n_{3})(m_{1}+ i m_{2}) -
(l_{0}-l_{3})(n_{0}-n_{3})(m_{1}- i m_{2}) \; ] \nonumber
\\
= l_{0} n_{3}m_{1} + l_{3} n_{0} m_{1} + i l_{0} n_{0} m_{2} + i
l_{3} n_{3} m_{2} \; ,
\nonumber
\end{eqnarray}
\begin{eqnarray}
{1 \over 2}\; [\;(m_{0} - m_{3}) ( n_{0} +n_{3}) (l_{1} + il_{2})
+ (m_{0} - m_{3}) ( n_{0} -n_{3}) (l_{1} - il_{2}) \; ] \nonumber
\\
= n_{0} m_{0} l_{1} - m_{3} n_{3} l_{1} + i m_{0} n_{3} l_{2} -
in_{0} m_{3} l_{2} \; , \nonumber
\\
{1 \over 2}\; [\;(m_{0} - m_{3}) ( n_{0} +n_{3}) (l_{1} + il_{2})
+ (m_{0} - m_{3}) ( n_{0} -n_{3}) (l_{1} - il_{2}) \; ] \nonumber
\\
= m_{0} n_{3} l_{1} - m_{3} n_{0} l_{1} + i m_{0} n_{0} l_{2} -
im_{3} n_{3} l_{2} \; ,
\nonumber
\end{eqnarray}
\begin{eqnarray}
-{1 \over 2}\; [\;  (m_{0} +m_{3})( l_{0} +l_{3}) (n_{1} + i
n_{2}) + (m_{0} -m_{3})( l_{0} -l_{3}) (n_{1} - i n_{2})  \; ]
\nonumber
\\
= -m_{0} l_{0} n_{1} - m_{3} l_{3} n_{1} - i m_{0} l_{3} n_{2} - i
m_{3} l_{0}n_{2} \; , \nonumber
\\
-{1 \over 2}\; [\;  (m_{0} +m_{3})( l_{0} +l_{3}) (n_{1} + i
n_{2}) - (m_{0} -m_{3})( l_{0} -l_{3}) (n_{1} - i n_{2})  \; ]
\nonumber
\\
= -m_{0} l_{3} n_{1} - m_{3} l_{0} n_{1} - i m_{0} l_{0} n_{2} - i
m_{3} l_{3}n_{2} \; ,
\nonumber
\end{eqnarray}
\begin{eqnarray}
-{1 \over 2}\; [\;(l_{1} + il_{2})(n_{1} + i n_{2})( m_{1} -i
m_{2})  + (l_{1} - il_{2})(n_{1} - i n_{2})( m_{1} +i m_{2}) \; ]
\nonumber
\\
= -l_{1} n_{1} m_{1} -l_{1} n_{2} m_{2} - l_{2} m_{2} n_{1} +
l_{2} n_{2} m_{1} \; , \nonumber
\\
-{1 \over 2}\; [\;(l_{1} + il_{2})(n_{1} + i n_{2})( m_{1} -i
m_{2})  - (l_{1} - il_{2})(n_{1} - i n_{2})( m_{1} +i m_{2}) \; ]
\nonumber
\\
= +i l_{1} n_{1} m_{2} -i l_{1} m_{1} n_{2} -i l_{2} m_{1} n_{1}
-i l_{2} n_{2} m_{2} \; ,
\nonumber
\end{eqnarray}

\noindent we find  $(k_{1})^{-1}$  and $(k_{2})^{-1}$:
\begin{eqnarray}
(k_{1})^{-1} = (-1) \; \{  k_{1} \; (mm) \nonumber
\\
+ l_{0} n_{0}m_{1} + l_{3} n_{3} m_{1} + i l_{0} n_{3} m_{2} + i
l_{3} n_{0} m_{2} \nonumber
\\
+n_{0} m_{0} l_{1} - m_{3} n_{3} l_{1} + i m_{0} n_{3} l_{2} -
in_{0} m_{3} l_{2} - \nonumber
\\
-m_{0} l_{0} n_{1} - m_{3} l_{3} n_{1} - i m_{0} l_{3} n_{2} - i
m_{3} l_{0}n_{2} \nonumber
\\
-l_{1} n_{1} m_{1} -l_{1} n_{2} m_{2} - l_{2} m_{2} n_{1} + l_{2}
n_{2} m_{1} \; \}  ,
\nonumber
\end{eqnarray}
\begin{eqnarray}
i(k_{2})^{-1} = (-1) \; \{  i k_{2} \; (mm) \nonumber
\\
+ l_{0} n_{3}m_{1} + l_{3} n_{0} m_{1} + i l_{0} n_{0} m_{2} + i
l_{3} n_{3} m_{2} \nonumber
\\
+ m_{0} n_{3} l_{1} - m_{3} n_{0} l_{1} + i m_{0} n_{0} l_{2} -
im_{3} n_{3} l_{2} \nonumber
\\
-m_{0} l_{3} n_{1} - m_{3} l_{0} n_{1} - i m_{0} l_{0} n_{2} - i
m_{3} l_{3}n_{2} \nonumber
\\
+i l_{1} n_{1} m_{2} -i l_{1} m_{1} n_{2} -i l_{2} m_{1} n_{1} -i
l_{2} n_{2} m_{2} \; .
\nonumber
\end{eqnarray}

\noindent From this, after identical transformations, we arrive at
\begin{eqnarray}
(k_{1})^{-1} = -k_{1} \; (mm) - m_{1} \; (ln) - l_{1} \; (nm) +
 n_{1 } (lm)
+ 2 \; [ \; {\bf l} \times ({\bf n} \times {\bf m}) \; ]_{1}
\nonumber
\\
+ i\; [\; m_{0} ( {\bf n} \times {\bf l} )_{1}  +
 l_{0} ( {\bf n} \times {\bf m} )_{1} +
n_{0} ( {\bf l} \times {\bf m} )_{1}  \; ] \;, \label{B.11a}
\end{eqnarray}
\begin{eqnarray}
(k_{2})^{-1} = -k_{2} \; (mm) - m_{2} \; (ln) - l_{2} \; (nm)
 n_{2 } (lm)
+ 2 \; [ \; {\bf l} \times ({\bf n} \times {\bf m}) \; ]_{2}
\nonumber
\\
+ i\; [\; m_{0} ( {\bf n} \times {\bf l} )_{2}  +
 l_{0} ( {\bf n} \times {\bf m} )_{2} +
n_{0} ( {\bf l} \times {\bf m} )_{2}  \; ] \;. \label{B.11b}
\end{eqnarray}

Thus, parameter  $(k_{a})^{-1}$ is defined as follows:
\begin{eqnarray}
(k_{0})^{-1} = k_{0} \; (mm) + m_{0} \; (ln) + l_{0} \; (nm)
-n_{0} (lm) +
  i \;   {\bf l} \;({\bf m} \times {\bf n} ) \; ,
\nonumber
\\[2mm]
(k_{j})^{-1} = -k_{j} \; (mm) - m_{j} \; (ln) - l_{j} \; (nm) +
 n_{j } (lm)
\nonumber
\\
 + 2 \; [ \; {\bf l} \times ({\bf n} \times {\bf m}) \; ]_{j} +
 i\; [\; m_{0} ( {\bf n} \times {\bf l} )_{j}  +
 l_{0} ( {\bf n} \times {\bf m} )_{j} +
n_{0} ( {\bf l} \times {\bf m} )_{j}  \; ] \;. \label{B.12}
\end{eqnarray}

One  may expect to obtain similar formulas for quantities
 $m,l,n$.

Now let us calculate
\begin{eqnarray}
(n_{0})^{-1} = {1 \over 2 \mid G \mid } \; (A_{42} + A_{31}) \;
,\qquad (n_{3})^{-1} = {1 \over 2 \mid G \mid } \; (A_{42} -
A_{31}) \; . \label{B.13}
\end{eqnarray}

\noindent Cofactor  $A_{42}$ is
\begin{eqnarray}
A_{42} = \left | \begin{array}{rrr}
+(k_{0} + k_{3})      &  +(n_{0} - n_{3})    &  -(n_{1} - in_{2}) \\
+(k_{1} + ik_{2} )    & -(n_{1} + in_{2})   &  + (n_{0} + n_{3} ) \\
-(l_{0} + l_{3})     &  +(m_{0} - m_{3})    &  -(m_{1} - im_{2})
\end{array} \right |
= \nonumber
\\
= -(l_{0}+l_{3})\; (nn) +  (k_{0} + k_{3})(n_{1} + in_{2})(m_{1}
-i m_{2}) \nonumber
\\
- (m_{0} -m_{3})(k_{1} +ik_{2})(n_{1} - i n_{2}) \nonumber
\\
+ (n_{0} -n_{3}) (k_{1} +k_{2})(m_{1} -im_{2}) \nonumber
\\
- (k_{0} + k_{3})( m_{0} -m_{3})(n_{0} + n_{3}) \; . \label{B.14a}
\end{eqnarray}

\noindent Cofactor $A_{31}$ is
\begin{eqnarray}
A_{31}= \left | \begin{array}{rrr}
  +(k_{1} - ik_{2})  &  +(n_{0} - n_{3})    &  -(n_{1} - in_{2}) \\
 + (k_{0} - k_{3} )  & -(n_{1} + in_{2})   &  + (n_{0} + n_{3} ) \\
-(l_{0} - l_{3})  & -(m_{1} + im_{2})   &   +(m_{0} + m_{3})
\end{array} \right | =
\nonumber
\\
= -(l_{0}-l_{3})\; (nn) + (k_{0} - k_{3})(n_{1} - in_{2})(m_{1} +i
m_{2}) \nonumber
\\
- (m_{0} +m_{3})(k_{1} -ik_{2})(n_{1} + i n_{2}) \nonumber
\\
+ (n_{0} +n_{3}) (k_{1} -k_{2})(m_{1} +im_{2}) \nonumber
\\
- (k_{0} - k_{3})( m_{0} +m_{3})(n_{0} - n_{3}) \; . \label{B.14b}
\end{eqnarray}

\noindent With the use of relations:
\begin{eqnarray}
{1 \over 2} \; [\; (k_{0} + k_{3})(n_{1} + in_{2})(m_{1} -i m_{2})
+ (k_{0} - k_{3})(n_{1} - in_{2})(m_{1} +i m_{2}) \; ] \nonumber
\\
= k_{0} n_{1} m_{1} + k_{0} n_{2} m_{2} - i k_{3} n_{1} m_{2} + i
k_{3} n_{2} m_{1} \; , \nonumber
\\
{1 \over 2} \; [\; (k_{0} + k_{3})(n_{1} + in_{2})(m_{1} -i m_{2})
- (k_{0} - k_{3})(n_{1} - in_{2})(m_{1} +i m_{2}) \; ] \nonumber
\\
= -i k_{0} n_{1} m_{2} + i k_{0} n_{2} m_{1} + k_{3} n_{1} m_{1} +
k_{3} n_{2} m_{2} \; ,
\nonumber
\end{eqnarray}
\begin{eqnarray}
-{1 \over 2} \; [\; (m_{0} -m_{3})(k_{1} +ik_{2})(n_{1} - i n_{2})
+ (m_{0} +m_{3})(k_{1} -ik_{2})(n_{1} + i n_{2}) \; ] \nonumber
\\
= - m_{0} k_{1} n_{1} - m_{0} k_{2} n_{2} - i m_{3} k_{1} n_{2} +
i m_{3} k_{2} n_{1} \; , \nonumber
\\
-{1 \over 2} \; [\; (m_{0} -m_{3})(k_{1} +ik_{2})(n_{1} - i n_{2})
- (m_{0} +m_{3})(k_{1} -ik_{2})(n_{1} + i n_{2}) \; ] \nonumber
\\
= +i m_{0} k_{1} n_{2} - i m_{0} k_{2} n_{1} +  m_{3} k_{1} n_{1}
+  m_{3} k_{2} n_{2} \; ,
\nonumber
\end{eqnarray}
\begin{eqnarray}
{1 \over 2} \; [\; (n_{0} -n_{3}) (k_{1} +k_{2})(m_{1} -im_{2}) +
(n_{0} +n_{3}) (k_{1} -k_{2})(m_{1} +im_{2}) \; ] \nonumber
\\
= n_{0} k_{1} m_{1} + n_{0} k_{2} m_{2} + i n_{3} k_{1} m_{2} - i
n_{3} k_{2} m_{1} \; , \nonumber
\\
{1 \over 2} \; [\; (n_{0} -n_{3}) (k_{1} +k_{2})(m_{1} -im_{2}) -
(n_{0} +n_{3}) (k_{1} -k_{2})(m_{1} +im_{2}) \; ] \nonumber
\\
= -i n_{0} k_{1} m_{2} + in_{0} k_{2} m_{1} - n_{3} k_{1} m_{1} -
n_{3} k_{2} m_{2} \; ,
\nonumber
\end{eqnarray}
\begin{eqnarray}
-{1 \over 2} \; [\; (k_{0} + k_{3})( m_{0} -m_{3})(n_{0} + n_{3})
+ (k_{0} - k_{3})( m_{0} +m_{3})(n_{0} - n_{3}) \; ] \nonumber
\\
= - k_{0} m_{0} n_{0} + k_{0} m_{3} n_{3} - k_{3} m_{0} n_{3}  +
k_{3} m_{3} n_{0} \; , \nonumber
\\
-{1 \over 2} \; [\; (k_{0} + k_{3})( m_{0} -m_{3})(n_{0} + n_{3})
- (k_{0} - k_{3})( m_{0} +m_{3})(n_{0} - n_{3}) \; ] \nonumber
\\
= - k_{0} m_{0} n_{3} + k_{0} m_{3} n_{0} - k_{3} m_{0} n_{0}  +
k_{3} m_{3} n_{3} \; ,
\nonumber
\end{eqnarray}

\noindent we  arrive at (factor  $\mid G \mid^{-1}$ is omitted)
\begin{eqnarray}
(n_{0})^{-1} = -l_{0}  \; (nn) \nonumber
\\
+ k_{0} n_{1} m_{1} + k_{0} n_{2} m_{2} - i k_{3} n_{1} m_{2} + i
k_{3} n_{2} m_{1} \nonumber
\\
- m_{0} k_{1} n_{1} - m_{0} k_{2} n_{2} - i m_{3} k_{1} n_{2} + i
m_{3} k_{2} n_{1} \nonumber
\\
+ n_{0} k_{1} m_{1} + n_{0} k_{2} m_{2} + i n_{3} k_{1} m_{2} - i
n_{3} k_{2} m_{1} \nonumber
\\
- k_{0} m_{0} n_{0} + k_{0} m_{3} n_{3} - k_{3} m_{0} n_{3}  +
k_{3} m_{3} n_{0}  \; , \nonumber
\end{eqnarray}
\begin{eqnarray}
(n_{3})^{-1} = -l_{3}  \; (nn) \nonumber
\\
-i k_{0} n_{1} m_{2} + i k_{0} n_{2} m_{1} + k_{3} n_{1} m_{1} +
k_{3} n_{2} m_{2} \nonumber
\\
 +i m_{0} k_{1} n_{2} - i m_{0} k_{2} n_{1} +  m_{3} k_{1} n_{1} +  m_{3} k_{2} n_{2}
\nonumber
\\
-i n_{0} k_{1} m_{2} + in_{0} k_{2} m_{1} - n_{3} k_{1} m_{1} -
n_{3} k_{2} m_{2} \nonumber
\\
- k_{0} m_{0} n_{3} + k_{0} m_{3} n_{0} - k_{3} m_{0} n_{0}  +
k_{3} m_{3} n_{3} \; .
\nonumber
\end{eqnarray}

\noindent From this it follows:
\begin{eqnarray}
(n_{0})^{-1} = - k_{0}\; (nm) + m_{0} \; (kn) - l_{0} \; (nn) -
n_{0} \; (km) \;+ \; i\; {\bf k} \; ({\bf m} \times {\bf n} ) \; ,
\nonumber
\\[2mm]
(n_{3})^{-1} = - k_{3}\; (nm) + m_{3} \; (kn) - l_{3} \; (nn) -
n_{3} \; (km) \; \nonumber
\\
+\; 2 \; [ \; {\bf k} \times ( {\bf m} \times {\bf n} ) \; ] _{3}
\;+  \;
 i k_{0} \;({\bf m} \times {\bf n}) _{3} +
i m_{0} \;({\bf k} \times {\bf n}) _{3} + i n_{0} \;({\bf m}
\times {\bf k}) _{3} \; . \label{B.17a}
\end{eqnarray}

Now, let us calculate
\begin{eqnarray}
-(n_{1})^{-1} = {A_{41} + A_{32} \over 2 \mid G \mid } \; , \qquad
i(n_{2})^{-1} = {A_{41} - A_{32} \over 2 \mid G \mid } \; .
\label{B.18}
\end{eqnarray}

\noindent Cofactor $A_{41}$  is
\begin{eqnarray}
A_{41} = (-1) \left | \begin{array}{ccc}
  +(k_{1} - ik_{2})  &  +(n_{0} - n_{3})    &  -(n_{1} - in_{2}) \\
 + (k_{0} - k_{3} )  & -(n_{1} + in_{2})   &  + (n_{0} + n_{3} ) \\
-(l_{1} - il_{2})  &  +(m_{0} - m_{3})    &  -(m_{1} - im_{2})
\end{array} \right |=
\nonumber
\\
= (-1) \{ \; - (l_{1} -il_{2}) \; (nn) +  (k_{1} -ik_{2})(n_{1}
+in_{2})(m_{1}-im_{2}) \nonumber
\\
- (k_{0}-k_{3})(m_{0}-m_{3})(n_{1} -in_{2})+ \nonumber
\\
+ (k_{0} -k_{3})(n_{0}-n_{3})(m_{1} -im_{2}) \nonumber
\\
- (m_{0} - m_{3})(n_{0} +n_{3})( k_{1} -ik_{2}) \; \}\;  ,
\label{B.19a}
\end{eqnarray}

\noindent Cofactor  $A_{32}$ is
\begin{eqnarray}
A_{32} = (-1) \left | \begin{array}{ccc}
+(k_{0} + k_{3})     &  +(n_{0} - n_{3})    &  -(n_{1} - in_{2}) \\
+(k_{1} + ik_{2} )   & -(n_{1} + in_{2})   &  + (n_{0} + n_{3} ) \\
-(l_{1} + il_{2})    & -(m_{1} + im_{2})   &   +(m_{0} + m_{3})
\end{array} \right | =
\nonumber
\\
= (-1) \; \{\; -(l_{1} + il_{2})\; (nn) +
 (k_{1} +ik_{2})(n_{1} -in_{2})(m_{1}+im_{2})
\nonumber
\\
- (k_{0}+k_{3})(m_{0}+m_{3})(n_{1} +in_{2}) \nonumber
\\
+ (k_{0} +k_{3})(n_{0}+n_{3})(m_{1} +im_{2}) \nonumber
\\
- (m_{0} + m_{3})(n_{0} -n_{3})( k_{1} +ik_{2}) \; \}\;  .
\label{B.19b}
\end{eqnarray}

\noindent Using the identities:
\begin{eqnarray}
-{1 \over 2} \; [\; (k_{1} -ik_{2})(n_{1} +in_{2})(m_{1}-im_{2}) +
(k_{1} +ik_{2})(n_{1} -in_{2})(m_{1}+im_{2}) \; ] \nonumber
\\
= -k_{1} n_{1} m_{1} - k_{1} n_{2} m_{2} + k_{2} n_{1} m_{2} -
k_{2} n_{2} m_{1} \; , \nonumber
\\
-{1 \over 2} \; [\; (k_{1} -ik_{2})(n_{1} +in_{2})(m_{1}-im_{2}) -
(k_{1} +ik_{2})(n_{1} -in_{2})(m_{1}+im_{2}) \; ] \nonumber
\\
= +i k_{1} n_{1} m_{2} -i k_{1} n_{2} m_{1} +i k_{2} n_{1} m_{1} +
i k_{2} n_{2} m_{2} \; ,
\nonumber
\end{eqnarray}
\begin{eqnarray}
{1 \over 2} \; [\; (k_{0}-k_{3})(m_{0}-m_{3})(n_{1} -in_{2}) +
(k_{0}+k_{3})(m_{0}+m_{3})(n_{1} +in_{2}) \; ] \nonumber
\\
= k_{0} m_{0} n_{1} + i k_{0} m_{3}n_{2} + ik_{3} m_{0} n_{2} +
k_{3} m_{3} n_{1} \; , \nonumber
\\
{1 \over 2} \; [\; (k_{0}-k_{3})(m_{0}-m_{3})(n_{1} -in_{2}) -
(k_{0}+k_{3})(m_{0}+m_{3})(n_{1} +in_{2}) \; ] \nonumber
\\
= -i k_{0} m_{0} n_{2} -  k_{0} m_{3}n_{1} - k_{3} m_{0} n_{1} -i
k_{3} m_{3} n_{2} \; ,
\nonumber
\end{eqnarray}
\begin{eqnarray}
-{1 \over 2} \; [\; (k_{0} -k_{3})(n_{0}-n_{3})(m_{1} -im_{2}) +
(k_{0} +k_{3})(n_{0}+n_{3})(m_{1} +im_{2}) \; ] \nonumber
\\
= - k_{0} n_{0} m_{1} -k_{3} n_{3} m_{1} -i k_{0} n_{3} m_{2} -i
k_{3} n_{0} m_{2}\; , \nonumber
\\
-{1 \over 2} \; [\; (k_{0} -k_{3})(n_{0}-n_{3})(m_{1} -im_{2}) -
(k_{0} +k_{3})(n_{0}+n_{3})(m_{1} +im_{2}) \; ] \nonumber
\\
= + k_{0} n_{3} m_{1} +k_{3} n_{0} m_{1} +i k_{0} n_{0} m_{2} +i
k_{3} n_{3} m_{2}\; ,
\nonumber
\end{eqnarray}
\begin{eqnarray}
{1 \over 2} \; [\; (m_{0} - m_{3})(n_{0} +n_{3})( k_{1} -ik_{2}) +
(m_{0} + m_{3})(n_{0} -n_{3})( k_{1} +ik_{2}) \; ] \nonumber
\\
= m_{0} n_{0} k_{1} - m_{3} n_{3} k_{1} - i m_{0} n_{3} k_{2} + i
m_{3} n_{0} k_{2} \;, \nonumber
\\
{1 \over 2} \; [\; (m_{0} - m_{3})(n_{0} +n_{3})( k_{1} -ik_{2}) -
(m_{0} + m_{3})(n_{0} -n_{3})( k_{1} +ik_{2}) \; ] \nonumber
\\
= m_{0} n_{3} k_{1} - m_{3} n_{0} k_{1} - i m_{0} n_{0} k_{2} + i
m_{3} n_{3} k_{2} \;,
\nonumber
\end{eqnarray}

\noindent we get expressions for  $(n_{1})^{-1}$  and
$(n_{2})^{-1}$:
\begin{eqnarray}
-(n_{1})^{-1} = l_{1} \; (nn) \nonumber
\\
-k_{1} n_{1} m_{1} - k_{1} n_{2} m_{2} + k_{2} n_{1} m_{2} - k_{2}
n_{2} m_{1} \nonumber
\\
+k_{0} m_{0} n_{1} + i k_{0} m_{3}n_{2} + ik_{3} m_{0} n_{2} +
k_{3} m_{3} n_{1} \nonumber
\\
- k_{0} n_{0} m_{1} -k_{3} n_{3} m_{1} -i k_{0} n_{3} m_{2} -i
k_{3} n_{0} m_{2} \nonumber
\\
+ m_{0} n_{0} k_{1} - m_{3} n_{3} k_{1} - i m_{0} n_{3} k_{2} + i
m_{3} n_{0} k_{2} \;,
\nonumber
\end{eqnarray}
\begin{eqnarray}
i(n_{2})^{-1} = -i l_{2} \; (nn) \nonumber
\\
+i k_{1} n_{1} m_{2} -i k_{1} n_{2} m_{1} +i k_{2} n_{1} m_{1} + i
k_{2} n_{2} m_{2} \nonumber
\\
-i k_{0} m_{0} n_{2} -  k_{0} m_{3}n_{1} - k_{3} m_{0} n_{1} -i
k_{3} m_{3} n_{2} \nonumber
\\
+ k_{0} n_{3} m_{1} +k_{3} n_{0} m_{1} +i k_{0} n_{0} m_{2} +i
k_{3} n_{3} m_{2} \nonumber
\\
+m_{0} n_{3} k_{1} - m_{3} n_{0} k_{1} - i m_{0} n_{0} k_{2} + i
m_{3} n_{3} k_{2}  \; .
\nonumber
\end{eqnarray}

\noindent From where, after identical transformations we arrive at
\begin{eqnarray}
(n_{1})^{-1} = - k_{1}\; (nm) + m_{1} \; (kn) - l_{1} \; (nn) -
n_{1} \; (km) \; \nonumber
\\
+\; 2 \; [ \; {\bf k} \times ( {\bf m} \times {\bf n} ) \; ] _{1}
\;+\;
 i k_{0} \;({\bf m} \times {\bf n}) _{1} +
i m_{0} \;({\bf k} \times {\bf n}) _{1} + i n_{0} \;({\bf m}
\times {\bf k}) _{2} \; ,
\nonumber
\\[2mm]
(n_{2})^{-1} = - k_{2}\; (nm) + m_{2} \; (kn) - l_{2} \; (nn) -
n_{2} \; (km) \; \nonumber
\\
+\; 2 \; [ \; {\bf k} \times ( {\bf m} \times {\bf n} ) \; ] _{2}
\;+ \; i k_{0} \;({\bf m} \times {\bf n}) _{2} + i m_{0} \;({\bf
k} \times {\bf n}) _{2} + i n_{0} \;({\bf m} \times {\bf k}) _{2}
\; . \label{B.22a}
\end{eqnarray}

Thus, parameter  $(n_{a})^{-1}$ is defined by
\begin{eqnarray}
(n_{0})^{-1} = - k_{0}\; (nm) + m_{0} \; (kn) - l_{0} \; (nn) -
n_{0} \; (km) \;+
 \; i\; {\bf k} \; ({\bf m} \times {\bf n} ) \; ,
\nonumber
\\[2mm]
(n_{j})^{-1} = - k_{j}\; (nm) + m_{j} \; (kn) - l_{j} \; (nn) -
n_{j} \; (km) \; \nonumber
\\
+\; 2 \; [ \; {\bf k} \times ( {\bf m} \times {\bf n} ) \; ] _{j}
\;+\;
 i k_{0} \;({\bf m} \times {\bf n}) _{j} +
i m_{0} \;({\bf k} \times {\bf n}) _{j} + i n_{0} \;({\bf m}
\times {\bf k}) _{j} \; . \label{B.23}
\end{eqnarray}

Let us calculate
\begin{eqnarray}
-(l_{0})^{-1} = {A_{13} + A_{24} \over 2 \mid G \mid } \;, \qquad
-(l_{3})^{-1} = {A_{13} - A_{24} \over 2 \mid G \mid } \; .
\label{B.24}
\end{eqnarray}

\noindent  Cofactor  $A_{13}$  is
\begin{eqnarray}
A_{13} = \left | \begin{array}{ccc}
+(k_{1} + ik_{2} )  & + (k_{0} - k_{3} )     &  + (n_{0} + n_{3} ) \\
-(l_{0} + l_{3})   & -(l_{1} - il_{2})      &  -(m_{1} - im_{2}) \\
-(l_{1} + il_{2})   & -(l_{0} - l_{3})     & +(m_{0} + m_{3})
\end{array} \right | =
\nonumber
\\
=  (n_{0} + n_{3}) \; (ll) - (m_{0} + m_{3})(k_{1} +ik_{2})(l_{1}
-il_{2}) \nonumber
\\
+ (k_{0} -k_{3})( m_{1} -im_{2})(l_{1} + i l_{2}) \nonumber
\\
+ (l_{0} +l_{3})( (k_{0} -k_{3})(m_{0} +m_{3}) \nonumber
\\
- (l_{0} - l_{3})( k_{1} + i k_{2})( m_{1} - i m_{2}) \; .
\label{B.25a}
\end{eqnarray}

\noindent Cofactor  $A_{24}$  is
\begin{eqnarray}
A_{24} = \left | \begin{array}{ccc}
+(k_{0} + k_{3})    &  +(k_{1} - ik_{2})  &  +(n_{0} - n_{3})     \\
-(l_{0} + l_{3})   & -(l_{1} - il_{2})  &  +(m_{0} - m_{3})     \\
-(l_{1} + il_{2})   & -(l_{0} - l_{3})  & -(m_{1} + im_{2})
\end{array} \right |=
\nonumber
\\
=  (n_{0} - n_{3}) \; (ll) -
 (m_{0} - m_{3})(k_{1} -ik_{2})(l_{1} +il_{2})
\nonumber
\\
+ (k_{0} + k_{3})( m_{1} + im_{2})(l_{1} - i l_{2}) + \nonumber
\\
+ (l_{0} - l_{3})( (k_{0} + k_{3})(m_{0} - m_{3}) \nonumber
\\
- (l_{0} + l_{3})( k_{1} - i k_{2})( m_{1} + i m_{2}) \; .
\label{B.25b}
\end{eqnarray}

\noindent Using the relations:
\begin{eqnarray}
-{1 \over 2} [\; (m_{0} + m_{3})(k_{1} +ik_{2})(l_{1} -il_{2}) +
(m_{0} - m_{3})(k_{1} -ik_{2})(l_{1} +il_{2})\; ] \nonumber
\\
=- m_{0} k_{1} l_{1} - m_{0} k_{2} l_{2} + i m_{3} k_{1} l_{2} - i
m_{3} k_{2}l_{1} \; , \nonumber
\\
-{1 \over 2} [\; (m_{0} + m_{3})(k_{1} +ik_{2})(l_{1} -il_{2}) -
(m_{0} - m_{3})(k_{1} -ik_{2})(l_{1} +il_{2}) \;] \nonumber
\\
= +i m_{0} k_{1} l_{2} - i m_{0} k_{2} l_{1} +  m_{3} k_{1} l_{1}
-  m_{3} k_{2}l_{2} \; ,
\nonumber
\end{eqnarray}
\begin{eqnarray}
{1 \over 2} \; [\; (k_{0} -k_{3})( m_{1} -im_{2})(l_{1} + i l_{2})
+ (k_{0} + k_{3})( m_{1} + im_{2})(l_{1} - i l_{2}) \; ] \nonumber
\\
= k_{0} m_{1} l_{1} + k_{0} m_{2} l_{2} - i k_{3} m_{1} l_{2} + i
k_{3} m_{2} l_{1} \; , \nonumber
\\
{1 \over 2} \; [\; (k_{0} -k_{3})( m_{1} -im_{2})(l_{1} + i l_{2})
- (k_{0} + k_{3})( m_{1} + im_{2})(l_{1} - i l_{2}) \; ] \nonumber
\\
= +i k_{0} m_{1} l_{2} -i k_{0} m_{2} l_{1} -  k_{3} m_{1} l_{1} -
k_{3} m_{2} l_{2} \; ,
\nonumber
\end{eqnarray}
\begin{eqnarray}
{1 \over 2 }[ \; (l_{0} +l_{3})( (k_{0} -k_{3})(m_{0} +m_{3}) +
(l_{0} - l_{3})( (k_{0} + k_{3})(m_{0} - m_{3})\; ] \nonumber
\\
= l_{0} k_{0}m_{0} -l_{0} k_{3} m_{3} + l_{3} k_{0}m_{3} -
l_{3}k_{3}m_{0} \; , \nonumber
\\
{1 \over 2 }[ \; (l_{0} +l_{3})( (k_{0} -k_{3})(m_{0} +m_{3}) -
(l_{0} - l_{3})( (k_{0} + k_{3})(m_{0} - m_{3})\; ] \nonumber
\\
= l_{0} k_{0}m_{3} -l_{0} k_{3} m_{0} + l_{3} k_{0}m_{0} -
l_{3}k_{3}m_{3} \; ,
\nonumber
\end{eqnarray}
\begin{eqnarray}
-{1 \over 2} [\; (l_{0} - l_{3})( k_{1} + i k_{2})( m_{1} - i
m_{2}) + (l_{0} + l_{3})( k_{1} - i k_{2})( m_{1} + i m_{2}) \; ]
\nonumber
\\
= -l_{0} k_{1} m_{1} - l_{0} k_{2} m_{2} - i l_{3} k_{1} m_{2} + i
l_{3} k_{2} m_{1} \; , \nonumber
\\
-{1 \over 2} [\; (l_{0} - l_{3})( k_{1} + i k_{2})( m_{1} - i
m_{2}) - (l_{0} + l_{3})( k_{1} - i k_{2})( m_{1} + i m_{2}) \; ]
\nonumber
\\
= +il_{0} k_{1} m_{2} - il_{0} k_{2} m_{1} +  l_{3} k_{1} m_{1} +
l_{3} k_{2} m_{2} \; ,
\nonumber
\end{eqnarray}

\noindent  we get
\begin{eqnarray}
-(l_{0})^{-1} =  + n_{0} \; (ll) \nonumber
\\
- m_{0} k_{1} l_{1} - m_{0} k_{2} l_{2} + i m_{3} k_{1} l_{2} - i
m_{3} k_{2}l_{1} \nonumber
\\
+ k_{0} m_{1} l_{1} + k_{0} m_{2} l_{2} - i k_{3} m_{1} l_{2} + i
k_{3} m_{2} l_{1} \nonumber
\\
+ l_{0} k_{0}m_{0} -l_{0} k_{3} m_{3} + l_{3} k_{0}m_{3} -
l_{3}k_{3}m_{0} \nonumber
\\
-l_{0} k_{1} m_{1} - l_{0} k_{2} m_{2} - i l_{3} k_{1} m_{2} + i
l_{3} k_{2} m_{1} \; ,
\nonumber
\end{eqnarray}
\begin{eqnarray}
-(l_{3})^{-1} =  +n_{3} \; (ll) \nonumber
\\
+i m_{0} k_{1} l_{2} - i m_{0} k_{2} l_{1} +  m_{3} k_{1} l_{1} -
m_{3} k_{2}l_{2} \nonumber
\\
+ i k_{0} m_{1} l_{2} -i k_{0} m_{2} l_{1} -  k_{3} m_{1} l_{1} -
k_{3} m_{2} l_{2} \nonumber
\\
+ l_{0} k_{0}m_{3} -l_{0} k_{3} m_{0} + l_{3} k_{0}m_{0} -
l_{3}k_{3}m_{3} \nonumber
\\
+ il_{0} k_{1} m_{2} - il_{0} k_{2} m_{1} +  l_{3} k_{1} m_{1} +
l_{3} k_{2} m_{2} \; .
\nonumber
\end{eqnarray}

\noindent From where we arrive at
\begin{eqnarray}
(l_{0})^{-1} = + k_{0} \; (ml) - m_{0} \; (kl) - l_{0} \; (km) -
n_{0}\; (ll) \;  +
 \; i\; {\bf m} \; ( {\bf l} \times {\bf k} )  \; ,
 \nonumber
 \\[2mm]
(l_{3})^{-1} = + k_{3} \; (ml) - m_{3} \; (kl) - l_{3} \; (km) -
n_{3}\; (ll) \nonumber
\\
+ 2\; [ {\bf m} \times ( {\bf k} \times {\bf l} )\; ] _{3} \; + \;
i\; m_{0} \; ({\bf l} \times {\bf k} ) _{3} + i\; k_{0} \; ({\bf
l} \times {\bf m} )_{3} + i\; l_{0} \; ({\bf m} \times {\bf
k})_{3} \; . \label{B.28b}
\end{eqnarray}

Now let us calculate
\begin{eqnarray}
-(l_{1})^{-1} = { A_{23} + A_{14} \over 2 \mid G \mid } \; ,
\qquad i(l_{2})^{-1} = { A_{23} - A_{14} \over 2 \mid G \mid } \;
. \label{B.29}
\end{eqnarray}

\noindent Cofactor  $A_{23}$  is
\begin{eqnarray}
A_{23} = (-1) \left | \begin{array}{ccc}
+(k_{0} + k_{3})    &  +(k_{1} - ik_{2})      &  -(n_{1} - in_{2}) \\
-(l_{0} + l_{3})   & -(l_{1} - il_{2})     &  -(m_{1} - im_{2}) \\
-(l_{1} + il_{2})   & -(l_{0} - l_{3})    & +(m_{0} + m_{3})
\end{array} \right |
= \nonumber
\\
(-1) \; [\; -(n_{1} -in_{2}) \; (ll) - (k_{0} +k_{3})( m_{0}
+m_{3})( l_{1} -il_{2}) \nonumber
\\
+ (l_{0} +l_{3})( m_{0} + m_{3})( k_{1} - ik_{2}) \nonumber
\\
- (l_{0} -l_{3})( k_{0} + k_{3})( m_{1} -im_{2}) \nonumber
\\
+ (k_{1} -ik_{2})( m_{1} -im_{2})(l_{1} + i l_{2}) \; ] \; ,
\label{B.30a}
\end{eqnarray}

\noindent Cofactor  $A_{14}$ is
\begin{eqnarray}
A_{14} = (-1) \left | \begin{array}{ccc}
+(k_{1} + ik_{2} )  & + (k_{0} - k_{3} )  & -(n_{1} + in_{2})    \\
-(l_{0} + l_{3})   & -(l_{1} - il_{2})  &  +(m_{0} - m_{3})     \\
-(l_{1} + il_{2})   & -(l_{0} - l_{3})  & -(m_{1} + im_{2})
\end{array} \right |=
\nonumber
\\
= (-1) \; [\; -(n_{1} +in_{2}) \; (ll) \nonumber
\\
- (k_{0} -k_{3})( m_{0} - m_{3})( l_{1} + il_{2}) \nonumber
\\
+ (l_{0} -l_{3})( m_{0} - m_{3})( k_{1} + ik_{2}) \nonumber
\\
- (l_{0} + l_{3})( k_{0} - k_{3})( m_{1} + i m_{2}) \nonumber
\\
+ (k_{1} +i k_{2})( m_{1} + i m_{2})(l_{1} - i l_{2}) \; ] \; .
\label{B.30b}
\end{eqnarray}

\noindent With the use of  identities:
\begin{eqnarray}
{1 \over 2} \;[\; (k_{0} +k_{3})( m_{0} +m_{3})( l_{1} -il_{2}) +
(k_{0} -k_{3})( m_{0} - m_{3})( l_{1} + il_{2}) \; ] \nonumber
\\
= l_{1} k_{0} m_{0} + l_{1} k_{3} m_{3} - i l_{2} k_{0} m_{3} - i
l_{2} k_{3} m_{0} \; , \nonumber
\\
{1 \over 2} \;[\; (k_{0} +k_{3})( m_{0} +m_{3})( l_{1} -il_{2}) -
(k_{0} -k_{3})( m_{0} - m_{3})( l_{1} + il_{2}) \; ] \nonumber
\\
= l_{1} k_{0} m_{3} + l_{1} k_{3} m_{0} - i l_{2} k_{0} m_{0} - i
l_{2} k_{3} m_{3} \; ,
\nonumber
\end{eqnarray}
\begin{eqnarray}
-{1 \over 2} \;[\; (l_{0} +l_{3})( m_{0} + m_{3})( k_{1} - ik_{2})
+ (l_{0} -l_{3})( m_{0} - m_{3})( k_{1} + ik_{2}) \; ] \nonumber
\\
= -  k_{1} l_{0}m_{0} -k_{1} l_{3} m_{3} + i k_{2} l_{0} m_{3} + i
k_{2} l_{3} m_{0} \; , \nonumber
\\
-{1 \over 2} \;[\; (l_{0} +l_{3})( m_{0} + m_{3})( k_{1} - ik_{2})
- (l_{0} -l_{3})( m_{0} - m_{3})( k_{1} + ik_{2}) \; ] \nonumber
\\
= -  k_{1} l_{0}m_{3} -k_{1} l_{3} m_{0} + i k_{2} l_{0} m_{0} + i
k_{2} l_{3} m_{3} \; ,
\nonumber
\end{eqnarray}
\begin{eqnarray}
{1 \over 2} \;[\; (l_{0} -l_{3})( k_{0} + k_{3})( m_{1} -im_{2}) +
(l_{0} + l_{3})( k_{0} - k_{3})( m_{1} + i m_{2}) \; ] \nonumber
\\
= m_{1} l_{0} k_{0} -m_{1} l_{3} k_{3} -im_{2} l_{0} k_{3} + i
m_{2} l_{3} k_{0} \; , \nonumber
\\
{1 \over 2} \;[\; (l_{0} -l_{3})( k_{0} + k_{3})( m_{1} -im_{2}) -
(l_{0} + l_{3})( k_{0} - k_{3})( m_{1} + i m_{2}) \; ] \nonumber
\\
= m_{1} l_{0} k_{3} -m_{1} l_{3} k_{0} -im_{2} l_{0} k_{0} + i
m_{2} l_{3} k_{3} \; ,
\nonumber
\end{eqnarray}
\begin{eqnarray}
-{1 \over 2} \;[\; (k_{1} -ik_{2})( m_{1} -im_{2})(l_{1} + i
l_{2})  + (k_{1} +i k_{2})( m_{1} + i m_{2})(l_{1} - i l_{2}) \; ]
\nonumber
\\
= - k_{1} m_{1} l_{1} - k_{1} m_{2} l_{2} - k_{2} m_{1} l_{2} +
k_{2} m_{2} l_{1} \; , \nonumber
\\
-{1 \over 2} \;[\; (k_{1} -ik_{2})( m_{1} -im_{2})(l_{1} + i
l_{2})  - (k_{1} +i k_{2})( m_{1} + i m_{2})(l_{1} - i l_{2}) \; ]
\nonumber
\\
= - i k_{1} m_{1} l_{2} + i  k_{1} m_{2} l_{1} + i  k_{2} m_{1}
l_{1} + i  k_{2} m_{2} l_{2} \; ,
\nonumber
\end{eqnarray}

\noindent  we get expressions for  $(l_{1})^{-1}$  and
$(l_{2})^{-1}$:
\begin{eqnarray}
- (l_{1})^{-1} = +  n_{1} \; (ll) \nonumber
\\
+ l_{1} k_{0} m_{0} + l_{1} k_{3} m_{3} - i l_{2} k_{0} m_{3} - i
l_{2} k_{3} m_{0}- \nonumber
\\
-  k_{1} l_{0}m_{0} -k_{1} l_{3} m_{3} + i k_{2} l_{0} m_{3} + i
k_{2} l_{3} m_{0} \nonumber
\\
+ m_{1} l_{0} k_{0} -m_{1} l_{3} k_{3} -im_{2} l_{0} k_{3} + i
m_{2} l_{3} k_{0} \nonumber
\\
- k_{1} m_{1} l_{1} - k_{1} m_{2} l_{2} - k_{2} m_{1} l_{2} +
k_{2} m_{2} l_{1} \; ,
\nonumber
\end{eqnarray}
\begin{eqnarray}
i\;  (l_{2})^{-1} = -i \;  n_{2} \; (ll) \nonumber
\\
+l_{1} k_{0} m_{3} + l_{1} k_{3} m_{0} - i l_{2} k_{0} m_{0} - i
l_{2} k_{3} m_{3} \nonumber
\\
-  k_{1} l_{0}m_{3} -k_{1} l_{3} m_{0} + i k_{2} l_{0} m_{0} + i
k_{2} l_{3} m_{3} \nonumber
\\
+ m_{1} l_{0} k_{3} -m_{1} l_{3} k_{0} -im_{2} l_{0} k_{0} + i
m_{2} l_{3} k_{3} \nonumber
\\
- i k_{1} m_{1} l_{2} + i  k_{1} m_{2} l_{1} + i  k_{2} m_{1}
l_{1} + i  k_{2} m_{2} l_{2} \; .
\nonumber
\end{eqnarray}

\noindent From where it follows
\begin{eqnarray}
(l_{1})^{-1} = + k_{1} \; (ml) - m_{1} \; (kl) - l_{1} \; (km) -
n_{1}\; (ll) \nonumber
\\
+ 2\; [ {\bf m} \times ( {\bf k} \times {\bf l} )\; ] _{1} \; + \;
i\; m_{0} \; ({\bf l} \times {\bf k} ) _{1} + i\; k_{0} \; ({\bf
l} \times {\bf m} )_{1} + i\; l_{0} \; ({\bf m} \times {\bf
k})_{1} \; ,
\nonumber
\\[2mm]
(l_{2})^{-1} = + k_{2} \; (ml) - m_{2} \; (kl) - l_{2} \; (km) -
n_{3}\; (ll) \nonumber
\\
+ 2\; [ {\bf m} \times ( {\bf k} \times {\bf l} )\; ] _{2} \; + \;
i\; m_{0} \; ({\bf l} \times {\bf k} ) _{2} + i\; k_{0} \; ({\bf
l} \times {\bf m} )_{2} + i\; l_{0} \; ({\bf m} \times {\bf
k})_{2} \; . \label{B.33b}
\end{eqnarray}

\noindent Thus, parameter   $(l)^{-1}$  is defined by
\begin{eqnarray}
(l_{0})^{-1} = + k_{0} \; (ml) - m_{0} \; (kl) - l_{0} \; (km) -
n_{0}\; (ll) \;  +
 \; i\; {\bf m} \; ( {\bf l} \times {\bf k} )  \; ,
 \nonumber
 \\[2mm]
(l_{j})^{-1} = + k_{j} \; (ml) - m_{j} \; (kl) - l_{3} \; (km) -
n_{j}\; (ll) \nonumber
\\
+ 2\; [ {\bf m} \times ( {\bf k} \times {\bf l} )\; ] _{j} \; + \;
i\; m_{0} \; ({\bf l} \times {\bf k} ) _{j} + i\; k_{0} \; ({\bf
l} \times {\bf m} )_{j} + i\; l_{0} \; ({\bf m} \times {\bf
k})_{j} \; . \label{B.34b}
\end{eqnarray}

It remains to calculate parameter  $(m)^{-1}$. For  $
(m_{0})^{-1}$  and $(m_{3})^{-1}$ we have
\begin{eqnarray}
(m_{0})^{-1} ={  A_{44}  + A_{33} \over 2 \mid G \mid } \; ,
\qquad (m_{3})^{-1} ={  A_{44}  - A_{33} \over 2 \mid G \mid }\; .
\label{B.35}
\end{eqnarray}

\noindent Cofactor  $A_{44}$ is
\begin{eqnarray}
A_{44} = \left | \begin{array}{ccc}
+(k_{0} + k_{3})    &  +(k_{1} - ik_{2})  &  +(n_{0} - n_{3})     \\
+(k_{1} + ik_{2} )  & + (k_{0} - k_{3} )  & -(n_{1} + in_{2})    \\
-(l_{0} + l_{3})   & -(l_{1} - il_{2})  &  +(m_{0} - m_{3})
\end{array} \right | =
\nonumber
\\
= (m_{0} -m_{3})\;  (kk) -
 (n_{0} -n_{3})( k_{1} + i k_{2}) (l_{1} - i l_{2})
\nonumber
\\
+ (l_{0} + l_{3})( k_{1} -i k_{2})( n_{1} + i n_{2}) \nonumber
\\
+ (l_{0} +l_{3})( (k_{0} -k_{3})( n_{0} -n_{3}) \nonumber
\\
- (k_{0} +k_{3})( l_{1} -i l_{2})( n_{1} + i n_{2}) \; ,
\label{B.36a}
\end{eqnarray}

\noindent Cofactor  $A_{33}$ is
\begin{eqnarray}
A_{33}= \left | \begin{array}{cccc}
+(k_{0} + k_{3})    &  +(k_{1} - ik_{2})     &  -(n_{1} - in_{2}) \\
+(k_{1} + ik_{2} )  & + (k_{0} - k_{3} )    &  + (n_{0} + n_{3} ) \\
-(l_{1} + il_{2})   & -(l_{0} - l_{3})     & +(m_{0} + m_{3})
\end{array} \right | =
\nonumber
\\
= (m_{0} +m_{3})\;  (kk) -
 (n_{0} +n_{3})( k_{1} - i k_{2}) (l_{1} + i l_{2})
\nonumber
\\
+ (l_{0} - l_{3})( k_{1} +i k_{2})( n_{1} - i n_{2}) \nonumber
\\
+ (l_{0} -l_{3})( (k_{0} + k_{3})( n_{0} + n_{3}) \nonumber
\\
- (k_{0} - k_{3})( l_{1} + i l_{2})( n_{1} - i n_{2}) \; .
\label{B.36b}
\end{eqnarray}

\noindent With the help of identities
\begin{eqnarray}
-{1 \over 2} \; [\; (n_{0} -n_{3})( k_{1} + i k_{2}) (l_{1} - i
l_{2}) + (n_{0} +n_{3})( k_{1} - i k_{2}) (l_{1} + i l_{2}) \; ]
\nonumber
\\
= -n_{0} k_{1} l_{1} - n_{0} k_{2} l_{2} -i n_{3} k_{1} l_{2} +i
n_{3} k_{2} l_{1} \; , \nonumber
\\
-{1 \over 2} \; [\; (n_{0} -n_{3})( k_{1} + i k_{2}) (l_{1} - i
l_{2}) - (n_{0} +n_{3})( k_{1} - i k_{2}) (l_{1} + i l_{2}) \; ]
\nonumber
\\
= +i n_{0} k_{1} l_{2} - i n_{0} k_{2} l_{1} + n_{3} k_{1} l_{1} +
n_{3} k_{2} l_{2} \; ,
\nonumber
\end{eqnarray}
\begin{eqnarray}
{1 \over 2} \; [\; (l_{0} + l_{3})( k_{1} -i k_{2})( n_{1} + i
n_{2}) + (l_{0} - l_{3})( k_{1} +i k_{2})( n_{1} - i n_{2}) \; ]
\nonumber
\\
= l_{0} k_{1} n_{1} + l_{0} k_{2} n_{2} + i l_{3} k_{1} n_{2} - i
l_{3} k_{2} n_{1} \; , \nonumber
\\
{1 \over 2} \; [\; (l_{0} + l_{3})( k_{1} -i k_{2})( n_{1} + i
n_{2}) - (l_{0} - l_{3})( k_{1} +i k_{2})( n_{1} - i n_{2}) \; ]
\nonumber
\\
= il_{0} k_{1} n_{2} - i l_{0} k_{2} n_{1} + l_{3} k_{1} n_{1} +
l_{3} k_{2} n_{2} \; ,
\nonumber
\end{eqnarray}
\begin{eqnarray}
{1 \over 2} \; [\; (l_{0} +l_{3})( k_{0} -k_{3})( n_{0} -n_{3}) +
(l_{0} -l_{3})( k_{0} + k_{3})( n_{0} + n_{3}) \; ] \nonumber
\\
= l_{0} k_{0}n_{0} +l_{0}k_{3}n_{3} - l_{3} k_{0} n_{3} - l_{3}
k_{3} n_{0} \; , \nonumber
\\
{1 \over 2} \; [\; (l_{0} +l_{3})( k_{0} -k_{3})( n_{0} -n_{3}) -
(l_{0} -l_{3})( k_{0} + k_{3})( n_{0} + n_{3}) \; ] \nonumber
\\
= -l_{0} k_{0} n_{3} - l_{0} k_{3} n_{0} + l_{3} k_{0} n_{0} +
l_{3} k_{3} n_{3} \; ,
\nonumber
\end{eqnarray}
\begin{eqnarray}
-{1 \over 2} \; [\; (k_{0} +k_{3})( l_{1} -i l_{2})( n_{1} + i
n_{2}) + (k_{0} - k_{3})( l_{1} + i l_{2})( n_{1} - i n_{2}) \; ]
\nonumber
\\
= - k_{0} l_{1} n_{1} - k_{0} l_{2} n_{2}  - i k_{3} l_{1} n_{2} +
i k_{3} l_{2} n_{1} \; , \nonumber
\\
-{1 \over 2} \; [\; (k_{0} +k_{3})( l_{1} -i l_{2})( n_{1} + i
n_{2}) - (k_{0} - k_{3})( l_{1} + i l_{2})( n_{1} - i n_{2}) \; ]
\nonumber
\\
= - i k_{0} l_{1} n_{2} + i k_{0} l_{2} n_{1} - k_{3} l_{1} n_{1}
+ k_{3} l_{2} n_{2} \; .
\nonumber
\end{eqnarray}

\noindent we get expressions for   $(m_{0})^{-1}$  and
$(m_{3})^{-1}$:
\begin{eqnarray}
(m_{0})^{-1}= + m_{0} \; (kk) \nonumber
\\
-n_{0} k_{1} l_{1} - n_{0} k_{2} l_{2} -i n_{3} k_{1} l_{2} +i
n_{3} k_{2} l_{1} \nonumber
\\
+ l_{0} k_{1} n_{1} + l_{0} k_{2} n_{2} + i l_{3} k_{1} n_{2} - i
l_{3} k_{2} n_{1} \nonumber
\\
+ l_{0} k_{0}n_{0} +l_{0}k_{3}n_{3} - l_{3} k_{0} n_{3} - l_{3}
k_{3} n_{0} \nonumber
\\
- k_{0} l_{1} n_{1} - k_{0} l_{2} n_{2}  - i k_{3} l_{1} n_{2} + i
k_{3} l_{2} n_{1} \; ,
\nonumber
\end{eqnarray}
\begin{eqnarray}
(m_{3})^{-1}= - m_{3} \; (kk) \nonumber
\\
+i n_{0} k_{1} l_{2} - i n_{0} k_{2} l_{1} + n_{3} k_{1} l_{1} +
n_{3} k_{2} l_{2} \nonumber
\\
+ il_{0} k_{1} n_{2} - i l_{0} k_{2} n_{1} + l_{3} k_{1} n_{1} +
l_{3} k_{2} n_{2} \nonumber
\\
-l_{0} k_{0} n_{3} - l_{0} k_{3} n_{0} + l_{3} k_{0} n_{0} + l_{3}
k_{3} n_{3} \nonumber
\\
- i k_{0} l_{1} n_{2} + i k_{0} l_{2} n_{1} - k_{3} l_{1} n_{1} +
k_{3} l_{2} n_{2}  \; .
\nonumber
\end{eqnarray}

\noindent From where we arrive at
\begin{eqnarray}
(m_{0})^{-1} = k_{0} \; (ln) + m_{0} \; (kk) - l_{0} (kn) + n_{0}
\; lk) \; +
 \;  i \; {\bf n} \; ({\bf l} \times {\bf k} ) \; ,
\nonumber
\\[2mm]
(m_{3})^{-1} =  - k_{3} \; (ln) - m_{3}\; (kk) + l_{3} \; (kn) -
n_{3}\; (kl) \; \nonumber
\\
+2\; [\; {\bf n} \times  ( {\bf l} \times {\bf k}) \; ]_{3}  \; +
\;  i\; n_{0} \; ({\bf k} \times {\bf l}) _{3} + i\; l_{0} \;(
{\bf k} \times {\bf n})_{3} + i\; k_{0} \; ({\bf n} \times {\bf
l})_{3} \; . \label{B.39b}
\end{eqnarray}

Now let us calculate  $(m_{1})^{-1}$  and $(m_{2})^{-1}$:
\begin{eqnarray}
-(m_{1})^{-1} = { A_{43} + A_{34} \over 2 \mid G \mid } \; ,
\qquad i(m_{2})^{-1} = { A_{43} - A_{34} \over 2 \mid G \mid } \;
. \label{B.40}
\end{eqnarray}

\noindent Cofactor  $A_{43} $  is
\begin{eqnarray}
A_{43} = (-1) \left | \begin{array}{ccc}
+(k_{0} + k_{3})    &  +(k_{1} - ik_{2})      &  -(n_{1} - in_{2}) \\
+(k_{1} + ik_{2} )  & + (k_{0} - k_{3} )  &  + (n_{0} + n_{3} ) \\
-(l_{0} + l_{3})   & -(l_{1} - il_{2})     &  -(m_{1} - im_{2})
\end{array} \right | =
\nonumber
\\
= (-1) \; [ \; -(m_{1} - i m_{2})\; (kk) + (k_{1} + i k_{2})(
n_{1} -i n_{2}) (l_{1} -i l_{2}) \; \nonumber
\\
- (k_{1} - i k_{2})( n_{0} + n_{3}) (l_{0} + l_{3}) \; \nonumber
\\
- (n_{1} - i n_{2})( l_{0} + l_{3})( k_{0} -k_{3}) \; \nonumber
\\
+ (l_{1} - il_{2} ) (k_{0} + k_{3}) (n_{0} + n_{3}) \; ] \; .
\label{B.41a}
\end{eqnarray}

\noindent Cofactor  $A_{43} $  is
\begin{eqnarray}
A_{34} = \left | \begin{array}{cccc}
+(k_{0} + k_{3})    &  +(k_{1} - ik_{2})  &  +(n_{0} - n_{3})     \\
+(k_{1} + ik_{2} )  & + (k_{0} - k_{3} )  & -(n_{1} + in_{2})    \\
-(l_{1} + il_{2})   & -(l_{0} - l_{3})  & -(m_{1} + im_{2})
\end{array} \right | =
\nonumber
\\
= (-1) \; [ \; -(m_{1} + i m_{2})\; (kk) + (k_{1} - i k_{2})(
n_{1} +i n_{2}) (l_{1} + i l_{2}) \; \nonumber
\\
- (k_{1} + i k_{2})( n_{0} - n_{3}) (l_{0} - l_{3}) \; \nonumber
\\
- (n_{1}  + i n_{2})( l_{0} - l_{3})( k_{0} + k_{3}) \; \nonumber
\\
+ (l_{1} + il_{2} ) (k_{0} - k_{3}) (n_{0} - n_{3}) \; ] \; .
\label{B.41b}
\end{eqnarray}

\noindent   With the use of relations:
\begin{eqnarray}
-{1 \over 2} \; [\; (k_{1} + i k_{2})( n_{1} -i n_{2}) (l_{1} -i
l_{2})  + (k_{1} - i k_{2})( n_{1} +i n_{2}) (l_{1} + i l_{2}) \;
] \nonumber
\\
= - k_{1} n_{1} l_{1} + k_{1} n_{2} l_{2} - k_{2} n_{1}l_{2} -
k_{2} n_{2} l_{1} \; , \nonumber
\\
-{1 \over 2} \; [\; (k_{1} + i k_{2})( n_{1} -i n_{2}) (l_{1} -i
l_{2})  - (k_{1} - i k_{2})( n_{1} +i n_{2}) (l_{1} + i l_{2}) \;
] \nonumber
\\
= +i k_{1} n_{1} l_{2} +i k_{1} n_{2} l_{1} -i  k_{2} n_{1}l_{1}
+i k_{2} n_{2} l_{2} \; ,
\nonumber
\end{eqnarray}
\begin{eqnarray}
{1 \over 2} \; [\; (k_{1} - i k_{2})( n_{0} + n_{3}) (l_{0} +
l_{3}) + (k_{1} + i k_{2})( n_{0} - n_{3}) (l_{0} - l_{3}) \; ]
\nonumber
\\
= k_{1} n_{0} l_{0}  + k_{1} n_{3} l_{3}  - i k_{2} n_{0} l_{3} -
i k_{2} n_{3} l_{0} \; , \nonumber
\\
{1 \over 2} \; [\; (k_{1} - i k_{2})( n_{0} + n_{3}) (l_{0} +
l_{3}) - (k_{1} + i k_{2})( n_{0} - n_{3}) (l_{0} - l_{3}) \; ]
\nonumber
\\
= k_{1} n_{0} l_{3}  + k_{1} n_{3} l_{0}  - i k_{2} n_{0} l_{0} -
i k_{2} n_{3} l_{3} \; ,
\nonumber
\end{eqnarray}
\begin{eqnarray}
{1 \over 2} \; [\; (n_{1} - i n_{2})( l_{0} + l_{3})( k_{0}
-k_{3})  + (n_{1}  + i n_{2})( l_{0} - l_{3})( k_{0} + k_{3}) \; ]
\nonumber
\\
= n_{1} l_{0} k_{0}  - n_{1} l_{3} k_{3} + i n_{2} l_{0} k_{3} - i
n_{2} l_{3} k_{0} \; , \nonumber
\\
{1 \over 2} \; [\; (n_{1} - i n_{2})( l_{0} + l_{3})( k_{0}
-k_{3})  - (n_{1}  + i n_{2})( l_{0} - l_{3})( k_{0} + k_{3}) \; ]
\nonumber
\\
= -n_{1} l_{0} k_{3}  + n_{1} l_{3} k_{0} - i n_{2} l_{0} k_{0} +
i n_{2} l_{3} k_{3} \; ,
\nonumber
\end{eqnarray}
\begin{eqnarray}
-{1 \over 2} \; [\; (l_{1} - il_{2} ) (k_{0} + k_{3}) (n_{0} +
n_{3}) + (l_{1} + il_{2} ) (k_{0} - k_{3}) (n_{0} - n_{3}) \; ]
\nonumber
\\
= - l_{1} k_{0} n_{0} - l_{1} k_{3} n_{3} + i l_{2} k_{0} n_{3} +
i l_{2} k_{3} n_{0} \; , \nonumber
\\
-{1 \over 2} \; [\; (l_{1} - il_{2} ) (k_{0} + k_{3}) (n_{0} +
n_{3}) - (l_{1} + il_{2} ) (k_{0} - k_{3}) (n_{0} - n_{3}) \; ]
\nonumber
\\
= - l_{1} k_{0} n_{3} - l_{1} k_{3} n_{0} + i l_{2} k_{0} n_{0} +
i l_{2} k_{3} n_{3} \; ,
\nonumber
\end{eqnarray}

\noindent we get
\begin{eqnarray}
-(m_{1})^{-1} = m_{1} \; (kk) \nonumber
\\
- k_{1} n_{1} l_{1} + k_{1} n_{2} l_{2} - k_{2} n_{1}l_{2} - k_{2}
n_{2} l_{1} \nonumber
\\
+ k_{1} n_{0} l_{0}  + k_{1} n_{3} l_{3}  - i k_{2} n_{0} l_{3} -
i k_{2} n_{3} l_{0} \nonumber
\\
+ n_{1} l_{0} k_{0}  - n_{1} l_{3} k_{3} + i n_{2} l_{0} k_{3} - i
n_{2} l_{3} k_{0} \nonumber
\\
- l_{1} k_{0} n_{0} - l_{1} k_{3} n_{3} + i l_{2} k_{0} n_{3} + i
l_{2} k_{3} n_{0} \; ,
\nonumber
\end{eqnarray}
\begin{eqnarray}
i(m_{2})^{-1} = -i \; m_{2} \; (kk) \nonumber
\\
+i k_{1} n_{1} l_{2} +i k_{1} n_{2} l_{1} -i  k_{2} n_{1}l_{1} +i
k_{2} n_{2} l_{2} \nonumber
\\
+ k_{1} n_{0} l_{3}  + k_{1} n_{3} l_{0}  - i k_{2} n_{0} l_{0} -
i k_{2} n_{3} l_{3} \nonumber
\\
-n_{1} l_{0} k_{3}  + n_{1} l_{3} k_{0} - i n_{2} l_{0} k_{0} + i
n_{2} l_{3} k_{3} \nonumber
\\
- l_{1} k_{0} n_{3} - l_{1} k_{3} n_{0} + i l_{2} k_{0} n_{0} + i
l_{2} k_{3} n_{3} \; .
\nonumber
\end{eqnarray}

\noindent From where we arrive at
\begin{eqnarray}
(m_{1})^{-1} =  - k_{1} \; (ln) - m_{1}\; (kk) + l_{1} \; (kn) -
n_{1}\; (kl) \; \nonumber
\\
+ \; 2\; [\; {\bf n} \times  ( {\bf l} \times {\bf k}) \; ]_{1} \;
+\;
 i\; n_{0} \; ({\bf k} \times {\bf l}) _{1} + i\; l_{0} \;( {\bf
k} \times {\bf n})_{1} + i\; k_{0} \; ({\bf n} \times {\bf l})_{1}
\; ,
\nonumber
\\[2mm]
(m_{2})^{-1} =  - k_{2} \; (ln) - m_{1}\; (kk) + l_{2} \; (kn) -
n_{2}\; (kl) \; \nonumber
\\
+ \; 2\; [\; {\bf n} \times  ( {\bf l} \times {\bf k}) \; ]_{2} \;
+ \;  i\; n_{0} \; ({\bf k} \times {\bf l}) _{2} + i\; l_{0} \;(
{\bf k} \times {\bf n})_{2} + i\; k_{0} \; ({\bf n} \times {\bf
l})_{2} \; . \label{B.44b}
\end{eqnarray}

Thus, the parameter  $(m)^{-1}$ is defined by
\begin{eqnarray}
(m_{0})^{-1} = k_{0} \; (ln) + m_{0} \; (kk) - l_{0} (kn) + n_{0}
\; lk) \; + \;  i \; {\bf n} \; ({\bf l} \times {\bf k} ) \; .
\nonumber
\\[2mm]
(m_{j})^{-1} =  - k_{j} \; (ln) - m_{j}\; (kk) + l_{j} \; (kn) -
n_{j}\; (kl) \; \nonumber
\\
+ 2\; [\; {\bf n} \times  ( {\bf l} \times {\bf k}) \; ]_{j}  \; +
 i\; n_{0} \; ({\bf k} \times {\bf l}) _{j} + i\; l_{0} \;( {\bf
k} \times {\bf n})_{j} + i\; k_{0} \; ({\bf n} \times {\bf l})_{j}
\; \label{B.45b}
\end{eqnarray}

Dirac parameters for the inverse matrix  $G^{-1}$ have been found;
it remains to determine determinant of $G$.

\section{ Determinant  $\mid G \mid $ in the Dirac parameters }

Collecting all results on parameters of the inverse matrix
$G^{-1}$ we have

\begin{eqnarray}
k_{0}' = \mid G \mid ^{-1} \; [\; k_{0} \; (mm) + m_{0} \; (ln) +
l_{0} \; (nm) - n_{0} (lm) +
  i \;   {\bf l} \;({\bf m} \times {\bf n} ) \; ] \;  , \;\;\;
\nonumber
\\
{\bf k} ' =  \mid G \mid ^{-1} \; [\; - {\bf k} \; (mm) - {\bf m}
\; (ln) - {\bf l} \; (nm) \; +
  {\bf n } (lm) \;+
  2 \; {\bf l} \times ({\bf n} \times {\bf m}) \;
\nonumber
\\
   +
i\;  m_{0} ( {\bf n} \times {\bf l} )  \;+ i\; l_{0} ( {\bf n}
\times {\bf m} ) + i\; n_{0} ( {\bf l} \times {\bf m} )  \; ] \;,
\nonumber
\\[2mm]
m_{0}'  = \mid G \mid ^{-1} \; [\;  k_{0} \; (ln) + m_{0} \; (kk)
- l_{0} (kn) + n_{0} \; lk) \; +
  i \; {\bf n} \; ({\bf l} \times {\bf k} ) \;] \;,\;\;
\nonumber
\\
{\bf m}' =  \mid G \mid ^{-1} \; [\; - {\bf k}  \; (ln) - {\bf
m}\; (kk) + {\bf l } \; (kn) - {\bf n}\; (kl) \; + \; 2\;  {\bf n}
\times  ( {\bf l} \times {\bf k})  \; \nonumber
\\
 +\;
 i\; n_{0} \; ({\bf k} \times {\bf l})  + i\; l_{0} \;( {\bf k} \times {\bf n}) +
i\; k_{0} \; ({\bf n} \times {\bf l}) \; ]\; ,
\nonumber
\end{eqnarray}
\begin{eqnarray}
l_{0}' = \mid G \mid ^{-1} \; [\; + k_{0} \; (ml) - m_{0} \; (kl)
- l_{0} \; (km) - n_{0}\; (ll) \;  + i\; {\bf m} \; ( {\bf l}
\times {\bf k} )  \; ]\; , \nonumber
\\
{\bf l} '  = \mid G \mid ^{-1} \; [\; + {\bf k} \; (ml) - {\bf m}
\; (kl) - {\bf l} \; (km) - {\bf n}\; (ll)\; +
  2\;  {\bf m} \times ( {\bf k} \times {\bf l} ) \;
\nonumber
\\
  + \;
 i\; m_{0} \; ({\bf l} \times {\bf k} )  + i\; k_{0} \; ({\bf l} \times {\bf m} ) +
i\; l_{0} \; ({\bf m} \times {\bf k}) \;]\;  , \nonumber
\\[2mm]
n_{0}' =  \mid G \mid ^{-1} \; [\;- k_{0}\; (nm) + m_{0} \; (kn) -
l_{0} \; (nn) - n_{0} \; (km) \;+
 \; i\; {\bf k} \; ({\bf m} \times {\bf n} ) \;] ,
\nonumber
\\
{\bf n}' =  \mid G \mid ^{-1} \; [\;- {\bf k}\; (nm) + {\bf m} \;
(kn)
 - {\bf l} \; (nn) - {\bf n} \; (km) \; +
\;   2 \;  {\bf k} \times ( {\bf m} \times {\bf n} ) \nonumber
\\
 +\;
 i k_{0} \; ({\bf m} \times {\bf n})  +
i m_{0} \;({\bf k} \times {\bf n})  + i n_{0} \;({\bf m} \times
{\bf k}) \; ] \; . \label{C.1}
\end{eqnarray}

\vspace{5mm} Let us substitute these expressions for inverse
parameters into determining relation  $G^{-1}$,  so we arrive at
the equations:
\begin{eqnarray}
1 = k_{0}'' = k_{0}' \; k_{0} + {\bf k}' \; {\bf k}
 - n'_{0}\;  l_{0} + {\bf n}' \; {\bf l} \;,
\label{C.2a}
\\
0= {\bf k}'' = k'_{0} \; {\bf k} + {\bf k}' \; k_{0}  + i \;  {\bf
k}' \times {\bf k} - n_{0}' \; {\bf l} + {\bf n}'\; l_{0} + i\;
{\bf n}' \times {\bf l} \; , \label{C.2b}
\\[2mm]
1 = m_{0}'' = m_{0}' \;  m_{0} + {\bf m}' \; {\bf m}
 - l'_{0}\;  n_{0} + {\bf l}' \; {\bf n} \;,
\label{C.3a}
\\
0= {\bf m}'' = m'_{0} \; {\bf m} + {\bf m}' \; m_{0}  - i \;  {\bf
m}' \times {\bf m} - l_{0}' \; {\bf n} + {\bf l}'\; n_{0} - i\;
{\bf l}' \times {\bf n} \; , \label{C.3b}
\\[2mm]
0= n_{0}'' = k_{0}' \; n_{0} - {\bf k}' \; {\bf n}
 + n'_{0} \; m_{0} + {\bf n}' \; {\bf m} \;,
\label{C.4a}
\\
0= {\bf n}'' = k'_{0} \; {\bf n} - {\bf k}' \; n_{0}  + i \;  {\bf
k}' \times {\bf n} + n_{0}' \; {\bf m} + {\bf n}'\; m_{0} - i\;
{\bf n}' \times {\bf m} \; , \label{C.4b}
\\[2mm]
0= l_{0}'' = l_{0}' \;  k_{0} + {\bf l}' \; {\bf k}
 + m'_{0}  \; l_{0} - {\bf m}' \; {\bf l} \;,
\label{C.5a}
\\
0= {\bf l}'' = l'_{0} \; {\bf k} + {\bf l}' \; k_{0}  + i \;  {\bf
l}' \times {\bf k} + m_{0}' \; {\bf l} - {\bf m}'\; l_{0} - i\;
{\bf m}' \times {\bf l } \; . \label{C.5b}
\end{eqnarray}

Consider eq. (\ref{C.2a}):
\begin{eqnarray}
1 = k_{0}' \; k_{0}
 - n'_{0}\;  l_{0} + {\bf k}' \; {\bf k} + {\bf n}' \; {\bf l} = ( \mbox{det} \; G )^{-1}
\nonumber
\\
\times \; [\; k_{0}^{2} (mm) + m_{0} k_{0} (ln) +
 l_{0}k_{0} (nm) -  n_{0} k_{0} (lm) + i k_{0} {\bf l}  ({\bf m} \times {\bf n})
\nonumber
\\
+ k_{0}l_{0} (nm) - m_{0} l_{0}  (kn) + l_{0}^{2} (nn) + n_{0}
l_{0} (km) - i l_{0} {\bf k} ({\bf m} \times {\bf n}) \nonumber
\\
- {\bf k}^{2}  (mm) - {\bf k} {\bf m} (ln)  - {\bf l} {\bf k} (nm)
+ {\bf n} {\bf k}  (lm) + 2 {\bf k} ({\bf l} \times ({\bf n}
\times {\bf m} ) ) \nonumber
\\
+ in_{0} {\bf k} ({\bf l} \times {\bf m}) + i l_{0} {\bf k} ({\bf
n} \times {\bf m}) + i m_{0} {\bf k} ({\bf n} \times {\bf l})
\nonumber
\\
- {\bf k} {\bf l} (nm) + {\bf m} {\bf l} (kn) -{\bf l}^{2} (nn) -
{\bf n} {\bf l} (km) + 2{\bf l} ({\bf k} \times ({\bf m} \times
{\bf n})) \nonumber
\\
+ i k_{0} {\bf l} ({\bf m} \times {\bf n}) + im_{0} {\bf l} ( {\bf
k} \times {\bf n}) + i n_{0} {\bf l} ({\bf m} \times {\bf k}) \;]
\;  . \nonumber
\end{eqnarray}

\noindent After  transformations it gives
\begin{eqnarray}
\mbox{det} \; G = (kk)\; (mm) + (ll) \; (nn) + 2\;  (mk) \; (ln)
+2 \; (lk)\; (nm) -2\; (nk)\; (lm) \nonumber
\\
+\;  2\; i \;[\; k_{0} \;  {\bf l}  ({\bf m} \times {\bf n}) \;+\;
m_{0} \;  {\bf k}  ({\bf n} \times {\bf l}) \;+\; l_{0} \;  {\bf
k}  ({\bf n} \times {\bf m}) \;+\; n_{0} \;  {\bf l}  ({\bf m}
\times {\bf k})   \; ] \; \nonumber
\\
+ \; 2 \;  {\bf k} \;  ({\bf l} \times ({\bf n} \times {\bf m} ) )
\; - \;2\;
 {\bf l} \;  ({\bf k} \times ({\bf n} \times {\bf m})) \;  .
\label{C.6a}
\end{eqnarray}

\noindent Taking in mind identity
\begin{eqnarray}
2 \; [\;\; {\bf k} \;  ({\bf l} \times ({\bf n} \times {\bf m} ) )
\; - \;
 {\bf l} \;  ({\bf k} \times ({\bf n} \times {\bf m})) \;\; ]=
  4 ({\bf k} {\bf n} ) \; ({\bf m} {\bf l}) -  4 ({\bf k} {\bf m} ) \; ({\bf n} {\bf l})
\nonumber
\end{eqnarray}

\noindent  for $\mbox{det} \;G$ we have
\begin{eqnarray}
\mbox{det} \; G = (kk)\; (mm) + (ll) \; (nn) + 2\;  (mk) \; (ln)
+2 \; (lk)\; (nm) -2\; (nk)\; (lm) \nonumber
\\
+\;  2\; i \;[\; k_{0} \;  {\bf l}  ({\bf m} \times {\bf n}) \;+\;
m_{0} \;  {\bf k}  ({\bf n} \times {\bf l}) \;+\; l_{0} \;  {\bf
k}  ({\bf n} \times {\bf m}) \;+\; n_{0} \;  {\bf l}  ({\bf m}
\times {\bf k})   \; ] \; \nonumber
\\
+ 4 ({\bf k} {\bf n} ) \; ({\bf m} {\bf l}) -  4 ({\bf k} {\bf m}
) \; ({\bf n} {\bf l})\; . \label{C.6b}
\end{eqnarray}

Now, let us verify that eq. (\ref{C.3a})  leads us to the same
$\mbox{det}\;G$. Indeed, from (\ref{C.3a}) it follows
\begin{eqnarray}
1 =  m_{0}' \;  m_{0} + {\bf m}' \; {\bf m}
 - l'_{0}\;  n_{0} + {\bf l}' \; {\bf n} = ( \mbox{det} \; G)^{-1} \;
\nonumber
\\
\times \; [\; k_{0} m_{0} \; (ln) + m_{0} ^{2} \; (kk) -
l_{0}m_{0} \; (kn) + n_{0}m_{0} \; (lk) + im_{0}\; {\bf n} ({\bf
l} \times {\bf k}) \nonumber
\\
- k_{0} n_{0} \; (ml) + m_{0}n_{0} \; (kl) + l_{0}n_{0} \; (km) +
n_{0}^{2} \; (ll) - in_{0}\; {\bf m} ({\bf l} \times {\bf k})
\nonumber
\\
- {\bf k}  {\bf m} \;(ln)  -{\bf m}^{2} \; (kk) + {\bf l}\;  {\bf
m} (kn) - {\bf n}  {\bf m}\; (lk) + 2{\bf m}\;  ({\bf n} \times
({\bf l} \times {\bf k})) \nonumber
\\
+ in_{0}  \; {\bf m}  ({\bf k} \times {\bf l}) + il_{0} \;  {\bf
m}  ({\bf k} \times {\bf n}) + i k_{0} \; {\bf m}  ({\bf n} \times
{\bf l}) \nonumber
\\
+ {\bf k} {\bf n} \; (ml)  -{\bf m} {\bf n}\; (kl)  -{\bf l}{\bf
n} \; (km)  -{\bf n}^{2}\; (ll) +2 {\bf n} \; ({\bf m} \times
({\bf k} \times {\bf l})) \nonumber
\\
+ im_{0}\; {\bf n} ({\bf l} \times {\bf k}) + ik_{0}\; {\bf n}
({\bf l} \times {\bf m}) + il_{0}\; {\bf n} ({\bf m} \times {\bf
k}) \; ] \; , \nonumber
\end{eqnarray}

\noindent from where we arrive at
\begin{eqnarray}
\mbox{det} \; G = (kk)\; (mm) + (ll) \; (nn) + 2\;  (mk) \; (ln)
+2 \; (lk)\; (nm) -2\; (nk)\; (lm) \nonumber
\\
+\;  2\; i \;[\; k_{0} \;  {\bf m}  ({\bf n} \times {\bf l}) \;+\;
m_{0} \;  {\bf n}  ({\bf l} \times {\bf k}) \;+\; l_{0} \;  {\bf
m}  ({\bf k} \times {\bf n}) \;+\; n_{0} \;  {\bf m}  ({\bf k}
\times {\bf l})   \; ] \; \nonumber
\\
+ \; 2 \;  {\bf m} \;  ({\bf n} \times ({\bf l} \times {\bf k} ) )
\; + \;2\;
 {\bf n} \;  ({\bf m} \times ({\bf k} \times {\bf l})) \;  .
\label{C.6c}
\end{eqnarray}

\noindent Taking in mind identity
\begin{eqnarray}
 2 \;  {\bf m} \;  ({\bf n} \times ({\bf l} \times {\bf k} ) ) \; + \;2\;
 {\bf n} \;  ({\bf m} \times ({\bf k} \times {\bf l})) =
4 ({\bf k} {\bf n} ) \; ({\bf m} {\bf l}) -  4 ({\bf k} {\bf m} )
\; ({\bf n} {\bf l})  \; ; \nonumber
\end{eqnarray}

\noindent  we see that eq. (\ref{C.6c}) coincides with
(\ref{C.6b}).

Now, let us turn to  relations  (\ref{C.4a}) and (\ref{C.5a}).
Firstly, consider eq. (\ref{C.4a}):
\begin{eqnarray}
0=  k_{0}' \; n_{0}  + n'_{0} \; m_{0}  - {\bf k}' \; {\bf n}
 + {\bf n}' \; {\bf m}
\nonumber
\\
= k_{0} n_{0} \; (m_{0}^{2} -{\bf m}^{2}) + m_{0} n_{0} \; (l_{0}
n_{0}  -{\bf l} {\bf n})+ l_{0} n_{0} \; (n_{0} m_{0}  -{\bf n}
{\bf m}) \nonumber
\\
- n_{0}^{2}  (l_{0} m_{0}  -{\bf l} {\bf m})+ in_{0} {\bf l} ({\bf
m} \times {\bf n}) \nonumber
\\
-k_{0} m_{0} \; (n_{0} m_{0} - {\bf n}{\bf m}) + m_{0}^{2} \;
(k_{0} n_{0}  -{\bf k} {\bf n})- l_{0} m_{0} \; (n_{0}^{2}  - {\bf
n}^{2}) \nonumber
\\
- n_{0}m_{0}  (k_{0} m_{0}  -{\bf k} {\bf m})+ im_{0} {\bf k}
({\bf m} \times {\bf n}) \nonumber
\\
+{\bf k}{\bf n}  \; (m_{0}^{2}  - {\bf m}^{2}) + {\bf m}{\bf n} \;
(l_{0} n_{0}  -{\bf l} {\bf n}) + {\bf l}{\bf n} \; (n_{0} m_{0}
-{\bf n} {\bf m}) - {\bf n}^{2}  \; (l_{0} m_{0}  -{\bf l} {\bf
m}) \nonumber
\\
- 2 {\bf n}\;  ({\bf l} \times ({\bf n} \times {\bf m})) - in_{0}
\; {\bf n} \; ({\bf l}\times {\bf m})  - il_{0} \;  {\bf n} \;
({\bf n} \times {\bf m}) - im_{0} \;  {\bf n} \; ({\bf n} \times
{\bf l}) \nonumber
\\
-{\bf k}{\bf m}  \; (n_{0}m_{0}   - {\bf n} {\bf m}) + {\bf m}^{2}
\; (k_{0} n_{0}  -{\bf k} {\bf n}) - {\bf l} {\bf m} \; (n_{0}^{2}
-{\bf n} ^{2}) - {\bf n}{\bf m}   \; (k_{0} m_{0}  -{\bf k} {\bf
m}) \nonumber
\\
+ 2 {\bf m}\;  ({\bf k} \times ({\bf m} \times {\bf n})) + ik_{0}
\; {\bf m} \; ({\bf m}\times {\bf n})  + im_{0} \;  {\bf m} \;
({\bf k} \times {\bf n}) + in_{0} \;  {\bf m} \; ({\bf m} \times
{\bf k}) . \label{C.7a}
\end{eqnarray}

\noindent One may verify that all terms with zero-index cancel out
each other, so that we have
\begin{eqnarray}
0=   + 2{\bf n}^{2}\; ({\bf l} {\bf m}) + 2 \; ({\bf k}{\bf m}
\;({\bf n}{\bf m})  -
 2 (\; {\bf k}{\bf n})\; ({\bf m}^{2}) - 2 \;  ({\bf l}{\bf n})\; ({\bf n}{\bf m})
\nonumber
\\
- 2 \; {\bf n}\;  ({\bf l} \times ({\bf n} \times {\bf m})) + 2
\;{\bf m}\;  ({\bf k} \times ({\bf m} \times {\bf n})) \; .
\nonumber
\end{eqnarray}

\noindent The later is equivalent to the  identity $0=0$:

\begin{eqnarray}
+ 2{\bf n}^{2}\; ({\bf l} {\bf m}) + 2 \; ({\bf k}{\bf m} \;({\bf
n}{\bf m})  -
 2 (\; {\bf k}{\bf n})\; ({\bf m}^{2}) - 2 \;  ({\bf l}{\bf n})\; ({\bf n}{\bf m})
\nonumber
\\
-2 {\bf n} \; [ {\bf n} \; ({\bf l}{\bf m}) - {\bf m} \; ( {\bf
l}{\bf n}) ] +2
 {\bf m} \; [ {\bf m} \; ({\bf k}{\bf n}) - {\bf n} \; ( {\bf k}{\bf m}) ] \equiv 0 \; .
 \label{C.7b}
 \end{eqnarray}

In the same manner consider eq. (\ref{C.5a}):

\begin{eqnarray}
0=  l_{0}' \;  k_{0}  + m'_{0}  \; l_{0} + {\bf l}' \; {\bf k}
  - {\bf m}' \; {\bf l}
\nonumber
\\
= k_{0}^{2}  \; (m_{0}l_{0}  -{\bf m} {\bf l} ) - m_{0} k_{0} \;
(k_{0} l_{0}  -{\bf k} {\bf l}) - l_{0} k_{0} \; (k_{0} m_{0}
-{\bf k} {\bf m}) - n_{0}k_{0}  (l_{0}^{2}   -{\bf l}^{2})+
ikn_{0} {\bf m} ({\bf l} \times {\bf k }) \nonumber
\\
+k_{0} l_{0} \; (l_{0} n_{0} - {\bf l}{\bf n}) + m_{0} l_{0} \;
(k_{0}^{2}  -{\bf k}^{2})- l_{0} ^{2} \; (k_{0} n_{0}   - {\bf
k}{\bf n}) + n_{0}l_{0}  (l_{0} k_{0}  -{\bf l} {\bf k})+ il_{0}
{\bf n} ({\bf l} \times {\bf k}) \nonumber
\\
+{\bf k}^{2}   \; (m_{0}l_{0}  - {\bf m} {\bf l}) - {\bf m}{\bf k}
\; (k_{0} l_{0}  -{\bf k} {\bf l}) - {\bf l}{\bf k} \; (k_{0}
m_{0}  -{\bf k} {\bf m}) - {\bf n}{\bf k}   \; (l_{0} ^{2}  -{\bf
l} ^{2}) \nonumber
\\
+ 2 {\bf k}\;  ({\bf m} \times ({\bf k} \times {\bf l})) + im_{0}
\; {\bf k} \; ({\bf l}\times {\bf k})  + ik_{0} \;  {\bf k} \;
({\bf l} \times {\bf m}) + il_{0} \;  {\bf k} \; ({\bf m} \times
{\bf k}) \nonumber
\\
+{\bf k}{\bf l}  \; (l_{0}n_{0}   - {\bf l} {\bf n}) + {\bf m}{\bf
l}  \; (k_{0}^{2}   -{\bf k}^{2}) - {\bf l} {\bf l} \; (k_{0}n_{0}
-{\bf k} {\bf n}) + {\bf n}{\bf l}   \; (l_{0} k_{0}  -{\bf l}
{\bf k}) \nonumber
\\
- 2 {\bf l}\;  ({\bf n} \times ({\bf l} \times {\bf k})) - in_{0}
\; {\bf l} \; ({\bf k}\times {\bf l})  - il_{0} \;  {\bf l} \;
({\bf k} \times {\bf n}) - ik_{0} \;  {\bf l} \; ({\bf n} \times
{\bf l}) \; , \label{C.8a}
\end{eqnarray}

\noindent and further
\begin{eqnarray}
0=   - 2{\bf k}^{2}\; ({\bf m} {\bf l}) + 2 \; ({\bf m}{\bf k}
\;({\bf k}{\bf l})  +
 2 (\; {\bf n}{\bf k})\; ({\bf l}^{2}) - 2 \;  ({\bf k}{\bf l})\; ({\bf l}{\bf n})
\nonumber
\\
+ 2 \; {\bf k}\;  ({\bf m} \times ({\bf k} \times {\bf l})) - 2
\;{\bf l}\;  ({\bf n} \times ({\bf l} \times {\bf k})) \; ;
\nonumber
\end{eqnarray}

\noindent which is equivalent to the identity  $0=0$:
\begin{eqnarray}
- 2\; {\bf k}^{2}\; ({\bf m} {\bf l}) + 2 \; ({\bf m}{\bf k}
\;({\bf k}{\bf l})  +
 2 \; (\; {\bf n}{\bf k})\; ({\bf l}^{2}) - 2 \;  ({\bf k}{\bf l})\; ({\bf l}{\bf n})
\nonumber
\\
+2  \; {\bf k} \; [ \; {\bf k} \; ({\bf m}{\bf l}) - {\bf l} \; (
{\bf m}{\bf k}) \; ] -2 \;
 {\bf l} \; [ \; {\bf l} \; ({\bf n}{\bf k}) - {\bf k} \; ( {\bf n}{\bf l}) \; ] \equiv 0 \; .
\label{C.8b}
\end{eqnarray}

\vspace{5mm} Equations  (\ref{C.2b}),
(\ref{C.3b}),(\ref{C.4b}),(\ref{C.5b}) can be verified as well.

In  the end of this section let us write down expression for $
\mbox{det} \; G$:
\begin{eqnarray}
\mbox{det} \; G = (kk)\; (mm) + (ll) \; (nn) + 2\;  (mk) \; (ln)
+2 \; (lk)\; (nm) -2\; (nk)\; (lm)  +\nonumber
\\
+\;  2\; i \;[\; k_{0} \;  {\bf l}  ({\bf m} \times {\bf n}) \;+\;
m_{0} \;  {\bf k}  ({\bf n} \times {\bf l}) \;+\; l_{0} \;  {\bf
k}  ({\bf n} \times {\bf m}) \;+\; n_{0} \;  {\bf l}  ({\bf m}
\times {\bf k})   \; ] \; \nonumber
\\
+ 4 ({\bf k} {\bf n} ) \; ({\bf m} {\bf l}) -  4 ({\bf k} {\bf m}
) \; ({\bf n} {\bf l})\; . \label{C.10}
\end{eqnarray}

\noindent The expression become more simple in special cases.

\vspace{5mm} {\bf Variant  A } \vspace{5mm}

All component with 0-index  are real, all component with index
1,2,3 are imaginary. Performing the change
\begin{eqnarray}
{\bf k} \Longrightarrow i\;{\bf k} ,\qquad {\bf m} \Longrightarrow
i\;{\bf m} ,\qquad {\bf l} \Longrightarrow i\;{\bf l} ,\qquad {\bf
n} \Longrightarrow i\;{\bf n} ,
 \label{C.11a}
 \end{eqnarray}

\noindent from  (\ref{C.10}) we get  (the notation is used: $[ab]
=  a_{0}a_{0} + a_{1}a_{1} +  a_{2}a_{2}  + a_{3} a_{3} $):
\begin{eqnarray}
\mbox{det} \; G = [kk]\; [mm] + [ll] \; [nn] + 2\;  [mk] \; [ln]
+2 \; [lk]\; [nm] -2\; [nk]\; [lm]) \nonumber
\\
+\;  2\;  \;[\; k_{0} \;  {\bf l}  ({\bf m} \times {\bf n}) \;+\;
m_{0} \;  {\bf k}  ({\bf n} \times {\bf l}) \;+\; l_{0} \;  {\bf
k}  ({\bf n} \times {\bf m}) \;+\; n_{0} \;  {\bf l}  ({\bf m}
\times {\bf k})   \; ] \; \nonumber
\\
+ 4 ({\bf k} {\bf n} ) \; ({\bf m} {\bf l}) -  4 ({\bf k} {\bf m}
) \; ({\bf n} {\bf l})\;, \label{C.11b}
\end{eqnarray}

\noindent here all the quantities are real-valued.

\vspace{15mm}
\begin{center}
 {\bf Variant  B }
 \end{center}
Restrictions  imposed are
\begin{eqnarray}
m_{a} = k^{*}_{a}, \qquad l_{a} = n^{*}_{a}  , \label{C.12a}
\end{eqnarray}

\noindent and  from (\ref{C.10})  it follows
\begin{eqnarray}
\mbox{det} \; G = (kk)\; (kk) ^{*} + (nn)^{*} \; (nn) + 2\;
(k^{*}k) \; (n^{*}n)
 +2 \; (n^{*}k)\; (nk^{*}) -2\; (nk)\; (nk)^{*}
\nonumber
\\
+\;  2\; i \;[\; k_{0} \;  {\bf n}^{*}  ({\bf k}^{*} \times {\bf
n}) \;+\; k_{0}^{*} \;  {\bf k}  ({\bf n} \times {\bf n}^{*})
\;+\; n_{0}^{*} \;  {\bf k}  ({\bf n} \times {\bf k}^{*}) \;+\;
n_{0} \;  {\bf n}^{*}  ({\bf k}^{*} \times {\bf k})   \; ] \;
\nonumber
\\
+ 4 ({\bf k} {\bf n} ) \; ({\bf k}^{*} {\bf n}^{*}) -
 4 ({\bf k} {\bf k}^{*} ) \; ({\bf n} {\bf n}^{*})\; .
\label{C.12b}
\end{eqnarray}

\noindent The latter can be rewritten in the form
\begin{eqnarray}
\mbox{det} \; G = (kk)\; (kk) ^{*} + (nn)^{*} \; (nn) + 2\;
(k^{*}k) \; (n^{*}n)
 +2 \; (n^{*}k)\; (nk^{*}) -2\; (nk)\; (nk)^{*}
\nonumber
\\
+\;  2\; i \;[\; k_{0} \;  {\bf k}^{*}  ({\bf n} \times {\bf
n}^{*}) \;-\; k_{0}^{*} \;  {\bf k}  ({\bf n}^{*} \times {\bf n})
\;+\; n_{0}^{*} \;  {\bf n}  ({\bf k} \times {\bf k}^{*}) \;-\;
n_{0} \;  {\bf n}^{*}  ({\bf k}^{*} \times {\bf k})   \; ] \;
\nonumber
\\
+ 4 ({\bf k} {\bf n} ) \; ({\bf k}^{*} {\bf n}^{*}) -
 4 ({\bf k} {\bf k}^{*} ) \; ({\bf n} {\bf n}^{*})\; ,
\label{C.12c}
\end{eqnarray}

\noindent which is explicitly real-valued.

\begin{center}
 {\bf  Variant C }
 \end{center}

In formulas  (\ref{C.11b}) one should take additional
restrictions:
\begin{eqnarray}
m_{0} = + k_{0}, \qquad  l_{0} = n_{0} , \qquad {\bf m} = - {\bf
k} , \qquad  {\bf l} = - {\bf n} \; ; \label{C.13a}
\\[2mm]
\mbox{det} \; G = [kk]\; [kk] + [nn] \; [nn] + 2\;  (kk) \; (nn)
+2 \; (nk)\; (nk) -2\; [nk]\; [nk]) \nonumber
\\
+\;  2\;  \;[\; k_{0} \;  {\bf n}  ({\bf k} \times {\bf n}) \;-\;
k_{0} \;  {\bf k}  ({\bf n} \times {\bf n}) \;-\; n_{0} \;  {\bf
k}  ({\bf n} \times {\bf k}) \;+\; n_{0} \;  {\bf n}  ({\bf k}
\times {\bf k})   \; ] \; \nonumber
\\
+ 4 ({\bf k} {\bf n} ) \; ({\bf k} {\bf n}) -  4 ({\bf k} {\bf k}
) \; ({\bf n} {\bf n}) \; ,
\nonumber
\end{eqnarray}

\noindent and further
\begin{eqnarray}
\mbox{det} \; G =
 [kk]\; [kk] + [nn] \; [nn]
\nonumber
\\
+ \; 2\;  (kk) \; (nn)  +2 \; (nk)\; (nk) -2\; [nk]\; [nk])  +
 4 ({\bf k} {\bf n} ) \; ({\bf k} {\bf n}) -  4 ({\bf k} {\bf k}
) \; ({\bf n} {\bf n})\; . \label{C.14c}
\end{eqnarray}

\section{ Independent calculation of $\mbox{det}\; G$}

Starting from the form
\begin{eqnarray}
G   =
 \left | \begin{array}{rrrr}
+(k_{0} + k_{3})    &  +(k_{1} - ik_{2})  &   +(n_{0} - n_{3})    &  -(n_{1} - in_{2}) \\
+(k_{1} + ik_{2} )  & + (k_{0} - k_{3} )  &  -(n_{1} + in_{2})   &  + (n_{0} + n_{3} ) \\
-(l_{0} + l_{3})   & -(l_{1} - il_{2})  &  +(m_{0} - m_{3})    &  -(m_{1} - im_{2}) \\
-(l_{1} + il_{2})   & -(l_{0} - l_{3})  &   -(m_{1} + im_{2})   &
+(m_{0} + m_{3})
\end{array} \right |
\label{D.1}
\end{eqnarray}

\noindent let us calculate $\mbox{det}\; G$ by direct method of
linear algebra:
\begin{eqnarray}
\mbox{det}\; G =  \left | \begin{array}{cccc}
G_{11} & G_{12} & G_{13} & G_{14} \\
G_{21} & G_{22} & G_{23} & G_{24} \\
G_{31} & G_{32} & G_{33} & G_{34} \\
G_{41} & G_{42} & G_{43} & G_{44}
\end{array} \right |
\nonumber
\\
= G_{11} \; \left | \begin{array}{ccc}
G_{22} & G_{23} & G_{24} \\
G_{32} & G_{33} & G_{34} \\
G_{42} & G_{43} & G_{44}
\end{array} \right | -
G_{12} \; \left | \begin{array}{ccc}
G_{21} & G_{23} & G_{24} \\
G_{31} & G_{33} & G_{34} \\
G_{41} & G_{43} & G_{44}
\end{array} \right | +
\nonumber
\\
+ G_{13} \; \left | \begin{array}{ccc}
G_{21} & G_{22} & G_{24} \\
G_{31} & G_{32} & G_{34} \\
G_{41} & G_{42} & G_{44}
\end{array} \right | -
G_{14} \; \left | \begin{array}{ccc}
G_{21} & G_{22} & G_{23} \\
G_{31} & G_{32} & G_{33} \\
G_{41} & G_{42} & G_{43}
\end{array} \right | \;
\nonumber
\end{eqnarray}

\noindent and  further
\begin{eqnarray}
\mbox{det}\; G \nonumber
\\
= G_{11} \;    ( \; G_{22} \left | \begin{array}{cc}
G_{33} & G_{34} \\
G_{43} & G_{44} \end{array} \right |
 - G_{23}
 \left | \begin{array}{cc}
G_{32} & G_{34} \\
G_{42} & G_{44} \end{array} \right | +
 G_{24}
 \left | \begin{array}{cc}
G_{32} & G_{33} \\
G_{42} & G_{43} \end{array} \right |  \;  ) \nonumber
\\
-G_{12} \;   ( \; G_{21} \left | \begin{array}{cc}
G_{33} & G_{34} \\
G_{43} & G_{44} \end{array} \right |
 - G_{23}
 \left | \begin{array}{cc}
G_{31} & G_{34} \\
G_{41} & G_{44} \end{array} \right | +
 G_{24}
 \left | \begin{array}{cc}
G_{31} & G_{33} \\
G_{41} & G_{43} \end{array} \right |  \;  ) \nonumber
\\
+G_{13} \;    ( \; G_{21} \left | \begin{array}{cc}
G_{32} & G_{34} \\
G_{42} & G_{44} \end{array} \right |
 - G_{22}
 \left | \begin{array}{cc}
G_{31} & G_{34} \\
G_{41} & G_{44} \end{array} \right | +
 G_{24}
 \left | \begin{array}{cc}
G_{31} & G_{32} \\
G_{41} & G_{42} \end{array} \right |  \;  ) \nonumber
\\
-G_{14} \;    ( \; G_{21} \left | \begin{array}{cc}
G_{32} & G_{33} \\
G_{42} & G_{43} \end{array} \right |
 - G_{22}
 \left | \begin{array}{cc}
G_{31} & G_{33} \\
G_{41} & G_{43} \end{array} \right | +
 G_{23}
 \left | \begin{array}{cc}
G_{31} & G_{32} \\
G_{41} & G_{42} \end{array} \right |  \;  ) \; . \nonumber
\end{eqnarray}

\noindent From this it follows
\begin{eqnarray}
\mbox{det}\; G = \nonumber
\\
= (G_{11}  \; G_{22} - G_{12} \; G_{21}  )\; \left |
\begin{array}{cc}
G_{33} & G_{34} \\
G_{43} & G_{44} \end{array} \right | + (- G_{11}  \; G_{23} +
G_{13} \; G_{21})\;
 \left | \begin{array}{cc}
G_{32} & G_{34} \\
G_{42} & G_{44} \end{array} \right | \nonumber
\\
+ (G_{11} \;  G_{24} -  G_{14} \; G_{21}  )\;
 \left | \begin{array}{cc}
G_{32} & G_{33} \\
G_{42} & G_{43} \end{array} \right |  + ( G_{12} \; G_{23}  -
G_{13} \; G_{22}  )\;
 \left | \begin{array}{cc}
G_{31} & G_{34} \\
G_{41} & G_{44} \end{array} \right | \nonumber
\\
+  ( -G_{12}  \; G_{24}   + G_{14} \; G_{22}  )\;
 \left | \begin{array}{cc}
G_{31} & G_{33} \\
G_{41} & G_{43} \end{array} \right | + (G_{13} \;  G_{24}  -
G_{14} \; G_{23}) \;
 \left | \begin{array}{cc}
G_{31} & G_{32} \\
G_{41} & G_{42} \end{array} \right | . \nonumber
\end{eqnarray}

\noindent Allowing for
\begin{eqnarray}
(G_{kl}) =  \left | \begin{array}{cccc}
+(k_{0} + k_{3})    &  +(k_{1} - ik_{2})   &  +(n_{0} - n_{3})     &   -(n_{1} - in_{2}) \\
+(k_{1} + ik_{2} )  &  + (k_{0} - k_{3} )  &  -(n_{1} + in_{2})    &   + (n_{0} + n_{3} ) \\
-(l_{0} + l_{3})    &  -(l_{1} - il_{2})   &  +(m_{0} - m_{3})     &   -(m_{1} - im_{2}) \\
-(l_{1} + il_{2})   &  -(l_{0} - l_{3})    &  -(m_{1} + im_{2}) &
+(m_{0} + m_{3})
\end{array} \right | \; ,
\nonumber
\end{eqnarray}

\noindent we arrive at the following form for $\mbox{det}\;G$:

\begin{eqnarray}
\mbox{det}\; G = \nonumber
\\
\; [ (k_{0} + k_{3})  (k_{0} - k_{3}) - (k_{1} - ik_{2}) (k_{1} +
ik_{2})  ] [ (m_{0} - m_{3})  (m_{0} + m_{3}) - (m_{1} - im_{2})
(m_{1} + im_{2}) ] \nonumber
\\
+ [  (k_{0} + k_{3})  (n_{1} + in_{2}) + (n_{0} - n_{3}) (k_{1} +
ik_{2} )  ] [ -(l_{1} - il_{2})  (m_{0} + m_{3}) - (m_{1} -
im_{2}) (l_{0} - l_{3})  ] \nonumber
\\
+[   (k_{0} + k_{3})  (n_{0} + n_{3} )  + (n_{1} - in_{2}) (k_{1}
+ ik_{2} ) ] [  (l_{1} - il_{2})  (m_{1} + im_{2}) + (m_{0} -
m_{3})  (l_{0} - l_{3})  ] \nonumber
\\
+[  -  (k_{1} - ik_{2})   (n_{1} + in_{2})  - (n_{0} - n_{3})
 (k_{0} - k_{3} ) ]  [  - (l_{0} + l_{3})  (m_{0} +
m_{3}) - (m_{1} - im_{2})(l_{1} + il_{2})  ] \nonumber
\\
+[   - (k_{1} - ik_{2})     (n_{0} + n_{3} )    - (n_{1} - in_{2})
(k_{0} - k_{3} )   ] [
 (l_{0} + l_{3}) (m_{1} + im_{2}) + (m_{0} - m_{3})  (l_{1} + il_{2}) ]
\nonumber
\\
+[ (n_{0} - n_{3})   (n_{0} + n_{3} )   - (n_{1} - in_{2}) (n_{1}
+ in_{2}) ] [  (l_{0} + l_{3}) (l_{0} - l_{3}) - (l_{1} -
il_{2})(l_{1} + il_{2})  ] \; . \label{D.2a}
\end{eqnarray}

\noindent and further (1-th and 6-th rows give simple terms)
\begin{eqnarray}
\mbox{det} G = ( k_{0}^{2} -  k_{1}^{2}  - k_{2}^{2} - k_{3}^{2} )
( m_{0}^{2} -  m_{1}^{2}  - m_{2}^{2} - m_{3}^{2} ) + ( n_{0}^{2}
-  n_{1}^{2}  - n_{2}^{2} - n_{3}^{2} ) ( l_{0}^{2} -  l_{1}^{2}
- l_{2}^{2} - l_{3}^{2} ) \nonumber
\\
+ [  (k_{0} + k_{3})  (n_{1} + in_{2}) + (n_{0} - n_{3}) (k_{1} +
ik_{2} )  ] [ -(l_{1} - il_{2})  (m_{0} + m_{3}) - (m_{1} -
im_{2}) (l_{0} - l_{3})  ] \nonumber
\\
+[   (k_{0} + k_{3})  (n_{0} + n_{3} )  + (n_{1} - in_{2}) (k_{1}
+ ik_{2} ) ] [ \; (l_{1} - il_{2})  (m_{1} + im_{2}) + (m_{0} -
m_{3})  (l_{0} - l_{3})  ] + \nonumber
\\
\;[  -  (k_{1} - ik_{2})   (n_{1} + in_{2})  - (n_{0} - n_{3})
 (k_{0} - k_{3} )\; ]  [  - (l_{0} + l_{3})  (m_{0} +
m_{3}) - (m_{1} - im_{2})(l_{1} + il_{2})  ] \nonumber
\\
+[   - (k_{1} - ik_{2})    (n_{0} + n_{3} )    - (n_{1} - in_{2})
(k_{0} - k_{3} )   ] [
 (l_{0} + l_{3}) (m_{1} + im_{2}) + (m_{0} - m_{3})  (l_{1} + il_{2}) ]  .
\label{D.2b}
\end{eqnarray}

Consider four separate terms:
\begin{eqnarray}
(1) = [ \; (k_{0} + k_{3}) \; (n_{1} + in_{2}) + (n_{0} - n_{3})
\; (k_{1} + ik_{2} ) \; ] \nonumber
\\
\times [\; -(l_{1} - il_{2}) \; (m_{0} + m_{3}) - (m_{1} -
im_{2})\; (l_{0} - l_{3})\;  ] \nonumber
\\
= [\;  (k_{0}n_{1} + k_{1} n_{0} ) + (  k_{3} n_{1} -  k_{1} n_{3}
) +
   i \; ( k_{0} n_{2}  +  k_{2} n_{0} ) + i \; ( k_{3} n_{2} - k_{2} n_{3}) \; ]
\nonumber
\\
\times [\; - (m_{0}l_{1} + m_{1}  l_{0})  +  ( m_{1} l_{3}  -
m_{3} l_{1} ) + i \; ( m_{0} l_{2} +  m_{2}  l_{0} )       + i \;
( m_{3} l_{2}  -  m_{2}  l_{3}) \;  ] \; ; \nonumber
\\[2mm]
(2) = [ \;  (k_{0} + k_{3})\;  (n_{0} + n_{3} )  + (n_{1} -
in_{2}) \; (k_{1} + ik_{2} )\; ]\; \nonumber
\\
  \times \;
[ \; (l_{1} - il_{2}) \; (m_{1} + im_{2}) + (m_{0} - m_{3}) \;
(l_{0} - l_{3})\;  ] \nonumber
\\
= [ \;     (k_{0}n_{3}  +k_{3}   n_{0} ) +   k_{0} n_{0} + ( n_{1}
k_{1} +n_{2} k_{2} + k_{3} n_{3} ) + i \; (n_{1}k_{2} -  n_{2}
k_{1} ) ] \; \nonumber
\\
\times \; [ \;  -(
    m_{0} l_{3} + m_{3} l_{0})  + m_{0} l_{0} + (l_{1} m_{1}
    + l_{2} m_{2} +m_{3} l_{3} ) \; +i \; ( l_{1} m_{2} - l_{2} m_{1})\;  ]\; ;
\nonumber
\\[2mm]
(3) = [ \; -  (k_{1} - ik_{2})  \; (n_{1} + in_{2})  - (n_{0} -
n_{3})  \; (k_{0} - k_{3} )\; ] \; \nonumber
\\
\times \; [\;  - (l_{0} + l_{3}) \; (m_{0} + m_{3}) - (m_{1} -
im_{2})(l_{1} + il_{2}) \; ] \nonumber
\\
\;[\;   ( n_{0} k_{3} + n_{3} k_{0})  - n_{0} k_{0}  - ( k_{1} n_{1}
+ k_{2} n_{2}  + n_{3} k_{3} )  + i \; ( k_{2} n_{1} - k_{1} n_{2}
) ] \; \nonumber
\\
\times \; [ \; -  ( l_{0} m_{3} + l_{3} m_{0} )  - l_{0} m_{0}  -
( l_{1} m_{1} + l_{2} m_{2} + l_{3} m_{3} ) + i\; ( l_{1} m_{2} -
l_{2} m_{1} ) \; ] \; ; \nonumber
\\[2mm]
(4) = [ \;  - (k_{1} - ik_{2})   \;  (n_{0} + n_{3} )    - (n_{1}
- in_{2})  \; (k_{0} - k_{3} )  \; ]\; \nonumber
\\
\times \; [\;
 (l_{0} + l_{3})\; (m_{1} + im_{2}) + (m_{0} - m_{3}) \; (l_{1} + il_{2})\; ]
\nonumber
\\
=[\; - ( k_{1} n_{0} + k_{0} n_{1})   + (n_{1} k_{3}  - k_{1}n_{3}
)  + i \; ( k_{2} n_{0} + n_{2} k_{0} ) + i\; (k_{2} n_{3} -n_{2}
k_{3} ) \; ] \; \nonumber
\\
\times \; [\;  ( l_{0} m_{1} +  l_{1} m_{0}  )     +  ( l_{3}
m_{1} - l_{1} m_{3} )
 + i \; (l_{0} m_{2} + l_{2}  m_{0} )   +  i \; ( l_{3} m_{2} - l_{2} m_{3} ) \; ] \; .
\nonumber
\end{eqnarray}

It is convenient to introduce the notation:
\begin{eqnarray}
{\bf k}  \times  {\bf n}  = {\bf A } \; , \qquad {\bf m}  \times
{\bf l}  = {\bf B } \; , \label{D.3}
\end{eqnarray}

\noindent then previous formulas look shorter
\begin{eqnarray}
\mbox{det}\; G =  (\; (kk) \; (mm) + (nn)\;(ll) + (1) + (2) + (3)
= (4) \; , \label{D.4} \nonumber
\\[2mm]
(1) = [\;  (k_{0}n_{1} + k_{1} n_{0} ) + A_{2}  +
   i \; ( k_{0} n_{2}  +  k_{2} n_{0} ) -  i \; A_{1} \; ]
\nonumber
\\
\times [\; - (m_{0}l_{1} + m_{1}  l_{0})  -B_{2}    + i \; ( m_{0}
l_{2} +  m_{2}  l_{0} )       - i B_{1}  \;  ] \; ,
\nonumber
\\[2mm]
(2) = [ \;   (  k_{0}n_{3}  +k_{3}   n_{0})  +   k_{0} n_{0} +
{\bf n} {\bf  k} -  i A_{3} \;  ] \; \nonumber
\\
\times \; [ \;  -(
    m_{0} l_{3} + m_{3} l_{0})  + m_{0} l_{0} + {\bf l} {\bf  m}
  -i B_{3} \;  ]\; ,
\nonumber
\\[2mm]
(3) = [\;   ( n_{0} k_{3} + n_{3} k_{0})  - n_{0} k_{0}  - {\bf k}
{\bf n}  - i \; A_{3} \;  ] \; \nonumber
\\
\times \; [ \; -l_{0} m_{0} -  ( l_{0} m_{3} + l_{3} m_{0} )  -
{\bf  l} {\bf m} - iB_{3} \;  ] \; ,
\nonumber
\\[2mm]
(4) = [\; - ( k_{1} n_{0} + k_{0} n_{1})   + A_{2}   + i \; (
k_{2} n_{0} + n_{2} k_{0} ) + iA_{1} \;
 ] \;
\nonumber
\\
\times \; [\;  ( l_{0} m_{1} +  l_{1} m_{0}  )     -B_{2}
 + i \; (l_{0} m_{2} + l_{2}  m_{0} )   +  i B_{1}  \; ] \; .
\nonumber
\end{eqnarray}

\noindent With the use of simplifying notation
\begin{eqnarray}
k_{0}\; n_{i}  = N_{i} \; , \qquad m_{0}\;  l_{i} = L_{i} \; ,
\qquad n_{0} \; k_{i} = K_{i} \; , \qquad l_{0} \; m_{i} = M_{i}
\; , \label{D.6}
\end{eqnarray}

\noindent  previous formulas look simpler:
\begin{eqnarray}
(1) = [\;  ( N_{1} + K_{1}  ) + A_{2}  +
   i \; ( N_{2}  +  K_{2}  ) -  i \; A_{1} \; ] \;
  \nonumber
  \\
  \times
    [\; - ( L_{1} + M_{1} )  -B_{2}    + i \;
( L_{2} +  M_{2}  )       - i B_{1}  \;  ] \; , \nonumber
\\
(2) = [ \;   (  N_{3}  + K_{3}  )  +   k_{0} n_{0} + {\bf n} {\bf
k} -  i A_{3} \;  ] \;
  \nonumber
  \\
  \times
[ \;  -(  L _{3} + M_{3} )  + m_{0} l_{0} +
{\bf l} {\bf  m}
  -i B_{3} \;  ]\; ;
\nonumber
\\
(3) = [\;   ( K_{3} + N_{3} )  - n_{0} k_{0}  - {\bf  k} {\bf n} -
i \; A_{3} \;  ] \;
  \nonumber
  \\
  \times
 [ \; -l_{0} m_{0} -  ( M_{3} + L_{3}  )  -
{\bf  l} {\bf m} - iB_{3} \;  ] \; ,
\nonumber
\\
(4) = [\; - ( K_{1}  + N_{1})   + A_{2}   + i \; ( K_{2} + N_{2} )
+ iA_{1} \;  ]
\nonumber
\\
\times  [\;  (  M_{1} +  L_{1}   )     -B_{2}
 + i \; ( M_{2} + L_{2}  )   +  i B_{1}  \; ] \; .
\label{D.7}\
\end{eqnarray}

\noindent Terms (1)-(4) in explicit form are
\begin{eqnarray}
(1) =
 - N_{1}  L_{1} -  N_{1} M_{1}   - N_{1} B_{2}    + i \; N_{1}
L_{2} +  i N_{1} M_{2}         - i N_{1} B_{1} \nonumber
\\
  -  K_{1} L_{1} - K_{1}  M_{1}   - K_{1} B_{2}    + i
K_{1}  L_{2} + i K_{1} M_{2}  - i  K_{1}B_{1} \nonumber
\\
  -  A_{2}  L_{1} - A_{2}  M_{1}   - A_{2} B_{2}    +
i  A_{2}  L_{2} + i A_{2} M_{2} - i  A_{2} B_{1} \nonumber
\\
 - i N_{2}  L_{1} -  i N_{2} M_{1}   - i N_{2} B_{2}    -
N_{2}  L_{2} -  N_{2}   M_{2}   +  N_{2} B_{1}  - \nonumber
\\
- i K_{2}  L_{1} -  i K_{2} M_{1}   - i K_{2} B_{2}    - K_{2}
L_{2} -  K_{2}   M_{2}   +  K_{2} B_{1} \nonumber
\\
 + i A_{1}  L_{1} +  i A_{1} M_{1}   + i A_{1} B_{2}    +
A_{1}  L_{2} +  A_{1}   M_{2}   -  A_{1} B_{1}  \; ,
\nonumber
\\[2mm]
(2) = - N_{3}    L _{3} - N_{3}  M_{3}   + N_{3} m_{0} l_{0} +
N_{3} {\bf l} {\bf  m}
  -i  N_{3} B_{3} \;
\nonumber
\\
 -  K_{3}   L _{3}  - K_{3}  M_{3}   + K_{3}   m_{0} l_{0} +  K_{3}{\bf l} {\bf  m}
  -i K_{3}  B_{3} \;-
\nonumber
\\
-  k_{0} n_{0}  L _{3} - k_{0} n_{0}  M_{3}   + k_{0} n_{0}  m_{0}
l_{0} + k_{0} n_{0}  {\bf l} {\bf  m}
  -i k_{0} n_{0}  B_{3} \;
\nonumber
\\
-  ({\bf n} {\bf  k})  L _{3} - ({\bf n} {\bf  k})   M_{3}   +
({\bf n} {\bf  k})   m_{0} l_{0} + ({\bf n} {\bf  k} ) ({\bf l}
{\bf  m})    -i ({\bf n} {\bf  k})  B_{3} \; \nonumber
\\
 +i A_{3}  L _{3} +  i A_{3} M_{3}    -  i A_{3} m_{0} l_{0} -  i A_{3} {\bf l} {\bf  m}
  - A_{3}  B_{3} \; ,
\nonumber
\end{eqnarray}
\begin{eqnarray}
(3) =
 - K_{3} l_{0} m_{0} -  K_{3}  M_{3} - K_{3}  L_{3}    -  K_{3} ({\bf  l} {\bf m}) -
i K_{3} B_{3} \; \nonumber
\\
- N_{3} l_{0} m_{0} -  N_{3}  M_{3} - N_{3}  L_{3}    - N_{3}({\bf
l} {\bf m}) - i N_{3} B_{3} \; \nonumber
\\
+  n_{0} k_{0}l_{0} m_{0}  + n_{0} k_{0}  M_{3} + n_{0} k_{0}
L_{3}   +   n_{0} k_{0} ( {\bf  l} {\bf m})
 + i  n_{0} k_{0} B_{3} \;
\nonumber
\\
+ ({\bf  k} {\bf n}) l_{0} m_{0} + ({\bf  k} {\bf n})  M_{3} +
({\bf  k} {\bf n}) L_{3}    +
 ({\bf  k} {\bf n}) ( {\bf  l} {\bf m}) + i ({\bf  k} {\bf n}) B_{3} \;
\nonumber
\\
+ i \; A_{3} l_{0} m_{0} +  i \; A_{3}  M_{3} +  i \; A_{3} L_{3}
+i \; A_{3}( {\bf  l} {\bf m}) -
 \; A_{3} B_{3} \;  ,
\nonumber
\\[2mm]
(4) = [\; - ( K_{1}  + N_{1})   + A_{2}   + i \; ( K_{2} + N_{2} )
+ iA_{1} \; \nonumber
\\
\times \;
 ] \;  [\;  (  M_{1} +  L_{1}   )     -B_{2}  + i \; ( M_{2} + L_{2}  )   +  i B_{1}  \; ]
\nonumber
\\
= -K_{1}     M_{1} - K_{1}  L_{1}        +  K_{1} B_{2}  - i K_{1}
M_{2} -  i  K_{1} L_{2}  - i K_{1} B_{1}  \; \nonumber
\\
-N_{1}  M_{1} - N_{1}  L_{1}   + N_{1} B_{2}  - i N_{1} M_{2} -
N_{1} i L_{2} -  i N_{1} B_{1}  \; \nonumber
\\
+ A_{2} M_{1} + A_{2}  L_{1}  - A_{2} B_{2}  + i  A_{2}   M_{2} +
i A_{2}  L_{2} +  i A_{2}  B_{1}  \; \nonumber
\\
+ iK_{2}  M_{1} +  iK_{2}  L_{1}        - iK_{2} B_{2}  - K_{2}
M_{2} - K_{2}L_{2}   - K_{2} B_{1} \nonumber
\\
+ iN_{2} M_{1} +  iN_{2}  L_{1}  - iN_{2} B_{2}  - N_{2}  M_{2} -
N_{2}L_{2}   - N_{2} B_{1} \nonumber
\\
+iA_{1} M_{1} +  iA_{1}  L_{1} - iA_{1}  B_{2} - A_{1} M_{2} -
A_{1}L_{2}    -A_{1} B_{1}  \; .
\nonumber
\end{eqnarray}

\noindent Summing these four relations:
\begin{eqnarray}
(1) + (2) + (3) + (4) \nonumber
\\
= - N_{1}  L_{1} -  N_{1} M_{1}   - N_{1} B_{2}    + i \; N_{1}
L_{2} +  i N_{1} M_{2}         - i N_{1} B_{1} \nonumber
\\
  -  K_{1} L_{1} - K_{1}  M_{1}   - K_{1} B_{2}    + i
K_{1}  L_{2} + i K_{1} M_{2}  - i  K_{1}B_{1} \nonumber
\\
 -  A_{2}  L_{1} - A_{2}  M_{1}   - A_{2} B_{2}    +
i  A_{2}  L_{2} + i A_{2} M_{2} - i  A_{2} B_{1} \nonumber
\\
- i N_{2}  L_{1} -  i N_{2} M_{1}   - i N_{2} B_{2}    - N_{2}
L_{2} -  N_{2}   M_{2}   +  N_{2} B_{1}  - \nonumber
\\
- i K_{2}  L_{1} -  i K_{2} M_{1}   - i K_{2} B_{2}    - K_{2}
L_{2} -  K_{2}   M_{2}   +  K_{2} B_{1} \nonumber
\\
+ i A_{1}  L_{1} +  i A_{1} M_{1}   + i A_{1} B_{2}    + A_{1}
L_{2} +  A_{1}   M_{2}   -  A_{1} B_{1} \nonumber
\\
- N_{3}    L _{3} - N_{3}  M_{3}   + N_{3} m_{0} l_{0} +  N_{3}
{\bf l} {\bf  m}
  -i  N_{3} B_{3}
\nonumber
\\
 -  K_{3}   L _{3}  - K_{3}  M_{3}   + K_{3}   m_{0} l_{0} +  K_{3}{\bf l} {\bf  m}
  -i K_{3}  B_{3}
\nonumber
\\
-  k_{0} n_{0}  L _{3} - k_{0} n_{0}  M_{3}   + k_{0} n_{0}  m_{0}
l_{0} + k_{0} n_{0}  {\bf l} {\bf  m}
  -i k_{0} n_{0}  B_{3}
\nonumber
\\
-  ({\bf n} {\bf  k})  L _{3} - ({\bf n} {\bf  k})   M_{3}   +
({\bf n} {\bf  k})   m_{0} l_{0} + ({\bf n} {\bf  k} ) ({\bf l}
{\bf  m})    -i ({\bf n} {\bf  k})  B_{3} \nonumber
\\
 +i A_{3}  L _{3} +  i A_{3} M_{3}    -  i A_{3} m_{0} l_{0} -  i A_{3} {\bf l} {\bf  m} \; -
  - A_{3}  B_{3}
\nonumber
\end{eqnarray}
\begin{eqnarray}
 - K_{3} l_{0} m_{0} -  K_{3}  M_{3} - K_{3}  L_{3}    -  K_{3} ({\bf  l} {\bf m}) -
i K_{3} B_{3} \nonumber
\\
- N_{3} l_{0} m_{0} -  N_{3}  M_{3} - N_{3}  L_{3}    - N_{3}({\bf
l} {\bf m}) - i N_{3} B_{3} \nonumber
\\
+  n_{0} k_{0}l_{0} m_{0}  + n_{0} k_{0}  M_{3} + n_{0} k_{0}
L_{3}   +   n_{0} k_{0} ( {\bf  l} {\bf m})
 + i  n_{0} k_{0} B_{3}
 \nonumber
\\
+ ({\bf  k} {\bf n}) l_{0} m_{0} + ({\bf  k} {\bf n})  M_{3} +
({\bf  k} {\bf n}) L_{3}    +
 ({\bf  k} {\bf n}) ( {\bf  l} {\bf m}) + i ({\bf  k} {\bf n}) B_{3} \; +
\nonumber
\\
+ i \; A_{3} l_{0} m_{0} +  i \; A_{3}  M_{3} +  i \; A_{3} L_{3}
+i \; A_{3}( {\bf  l} {\bf m}) -
 \; A_{3} B_{3}
\nonumber
\\
-K_{1}     M_{1} - K_{1}  L_{1}        +  K_{1} B_{2}  - i  K_{1}
M_{2} -  i  K_{1} L_{2}  - i K_{1} B_{1} \nonumber
\\
-N_{1}  M_{1} - N_{1}  L_{1}   + N_{1} B_{2}  - i N_{1} M_{2} -
N_{1} i L_{2} -  i N_{1} B_{1} \nonumber
\\
+ A_{2} M_{1} + A_{2}  L_{1}  - A_{2} B_{2}  + i  A_{2}   M_{2} +
i A_{2}  L_{2} +  i A_{2}  B_{1} \nonumber
\\
+ iK_{2}  M_{1} +  iK_{2}  L_{1}        - iK_{2} B_{2}  - K_{2}
M_{2} - K_{2}L_{2}   - K_{2} B_{1} \nonumber
\\
+ iN_{2} M_{1} +  iN_{2}  L_{1}  - iN_{2} B_{2}  - N_{2}  M_{2} -
N_{2}L_{2}   - N_{2} B_{1} \nonumber
\\
+iA_{1} M_{1} +  iA_{1}  L_{1} - iA_{1}  B_{2} - A_{1} M_{2} -
A_{1}L_{2}    -A_{1} B_{1}  \; .
\nonumber
\end{eqnarray}

After simple evident simplifications we get
\begin{eqnarray}
(1) + (2) + (3) + (4) \nonumber
\\
= -2 \;{\bf N}{\bf L} - 2 \;{\bf N} {\bf M} - 2 \;{\bf K}{\bf L}
-2\; {\bf K}{\bf M}  - 2 \;{\bf A} {\bf B} \nonumber
\\
- 2i \;{\bf N} {\bf B}  - 2i\; {\bf K} {\bf B}    + 2i \;{\bf A}
{\bf L} + 2i \;{\bf A} {\bf M} + \nonumber
\\
+ N_{3} m_{0} l_{0} +  N_{3} ({\bf l} {\bf  m}) \nonumber
\\
  + K_{3}   m_{0} l_{0} +  K_{3} ({\bf l} {\bf  m})
  \nonumber
\\
-  k_{0} n_{0}  L _{3} - k_{0} n_{0}  M_{3}   + k_{0} n_{0}  m_{0}
l_{0} + k_{0} n_{0}  {\bf l} {\bf  m}
  -i k_{0} n_{0}  B_{3}
\nonumber
\\
-  ({\bf n} {\bf  k})  L _{3} - ({\bf n} {\bf  k})   M_{3}   +
({\bf n} {\bf  k})   m_{0} l_{0} + ({\bf n} {\bf  k} ) ({\bf l}
{\bf  m})    -i ({\bf n} {\bf  k})  B_{3} \nonumber
\\
     -  i A_{3} m_{0} l_{0} -  i A_{3} ({\bf l} {\bf  m} ) -
\nonumber
\\
 - K_{3} l_{0} m_{0}   -  K_{3} ({\bf  l} {\bf m})
\nonumber
\\
- N_{3} l_{0} m_{0}   - N_{3}({\bf  l} {\bf m}) \nonumber
\\
+  n_{0} k_{0}l_{0} m_{0}  + n_{0} k_{0}  M_{3} + n_{0} k_{0}
L_{3}   +   n_{0} k_{0} ( {\bf  l} {\bf m})
 + i  n_{0} k_{0} B_{3}
\nonumber
\\
+ ({\bf  k} {\bf n}) l_{0} m_{0} + ({\bf  k} {\bf n})  M_{3} +
({\bf  k} {\bf n}) L_{3}    +
 ({\bf  k} {\bf n}) ( {\bf  l} {\bf m}) + i ({\bf  k} {\bf n}) B_{3}
\nonumber
\\
+ i \; A_{3} l_{0} m_{0}     +i \; A_{3}( {\bf  l} {\bf m}) \; .
\label{D.10}
\end{eqnarray}

Noting that all terms containing   $N_{3}, K_{3}, A_{3}$ cancel
out each other, so we get
\begin{eqnarray}
(1) + (2) + (3) + (4) \nonumber
\\
= -2 \;{\bf N}{\bf L} - 2 \;{\bf N} {\bf M} - 2 \;{\bf K}{\bf L}
-2\; {\bf K}{\bf M}  - 2 \;{\bf A} {\bf B} \nonumber
\\- 2i \;{\bf N} {\bf B}  - 2i\; {\bf K} {\bf B}    + 2i \;{\bf A}
{\bf L} + 2i \;{\bf A} {\bf M} \nonumber
\\
-  k_{0} n_{0}  L _{3} - k_{0} n_{0}  M_{3}   + k_{0} n_{0}  m_{0}
l_{0} + k_{0} n_{0}  {\bf l} {\bf  m}
  -i k_{0} n_{0}  B_{3}
\nonumber
\\
-  ({\bf n} {\bf  k})  L _{3} - ({\bf n} {\bf  k})   M_{3}   +
({\bf n} {\bf  k})   m_{0} l_{0} + ({\bf n} {\bf  k} ) ({\bf l}
{\bf  m})    -i ({\bf n} {\bf  k})  B_{3} \nonumber
\\
+  n_{0} k_{0}l_{0} m_{0}  + n_{0} k_{0}  M_{3} + n_{0} k_{0}
L_{3}   +   n_{0} k_{0} ( {\bf  l} {\bf m})
 + i  n_{0} k_{0} B_{3}
 \nonumber
\\
+ ({\bf  k} {\bf n}) l_{0} m_{0} + ({\bf  k} {\bf n})  M_{3} +
({\bf  k} {\bf n}) L_{3}    +
 ({\bf  k} {\bf n}) ( {\bf  l} {\bf m}) + i ({\bf  k} {\bf n}) B_{3} \; .
\label{D.11}
\end{eqnarray}

\noindent and further
\begin{eqnarray}
(1) + (2) + (3) + (4) \nonumber
\\
= -2 \;{\bf N}{\bf L} - 2 \;{\bf N} {\bf M} - 2 \;{\bf K}{\bf L}
-2\; {\bf K}{\bf M}  - 2 \;{\bf A} {\bf B} \nonumber
\\
- 2i \;{\bf N} {\bf B}  - 2i\; {\bf K} {\bf B}    + 2i \;{\bf A}
{\bf L} + 2i \;{\bf A} {\bf M} \nonumber
\\
   + 2 \;  k_{0} n_{0}  m_{0} l_{0} + 2 \;  k_{0} n_{0}  \; ({\bf l} {\bf  m})
 +  2 \; ({\bf n} {\bf  k})  \;  m_{0} l_{0} +
2 \; ({\bf n} {\bf  k} )\;  ({\bf l} {\bf  m})  \; . \label{D.12}
\end{eqnarray}

Therefore, determinant of $G$ is  given by
\begin{eqnarray}
\mbox{det}\; G
 (kk) \; (mm) + (nn)\;(ll) \;
\nonumber
\\
-\; 2 \;{\bf N}{\bf L} - 2 \;{\bf N} {\bf M} - 2 \;{\bf K}{\bf L}
-2\; {\bf K}{\bf M}  - 2 \;{\bf A} {\bf B} \nonumber
\\
- \; 2i \;{\bf N} {\bf B}  - 2i\; {\bf K} {\bf B}    + 2i \;{\bf
A} {\bf L} + 2i \;{\bf A} {\bf M} \; + \nonumber
\\
   + \; 2 \;  k_{0} n_{0}  (  m_{0} l_{0} + {\bf l} {\bf  m})
 +  2 \; ({\bf n} {\bf  k})  (  \;  m_{0} l_{0} +
{\bf l} {\bf  m})  \; , \label{D.13}
\end{eqnarray}

\noindent or by a shorter relation
\begin{eqnarray}
\mbox{det}\; G =
 (kk) \; (mm) + (nn)\;(ll)
\nonumber
\\
 + \; 2 \; (  k_{0} n_{0} + {\bf n} {\bf  k}   )\; (  m_{0} l_{0} + {\bf l} {\bf  m})
\nonumber
\\
-2 \;{\bf N}{\bf L} - 2 \;{\bf N} {\bf M} - 2 \;{\bf K}{\bf L}
-2\; {\bf K}{\bf M}  - 2 \;{\bf A} {\bf B} \nonumber
\\
- 2i \;{\bf N} {\bf B}  - 2i\; {\bf K} {\bf B}    + 2i \;{\bf A}
{\bf L} + 2i \;{\bf A} {\bf M} \;. \label{D.14a}
\end{eqnarray}

\noindent The latter formulas allows further simplification:
\begin{eqnarray}
\mbox{det}\; G
 (kk) \; (mm) + (nn)\;(ll)
\nonumber
\\
 + 2 \; (  k_{0} n_{0} + {\bf n} {\bf  k}   )\; (  m_{0} l_{0} + {\bf l} {\bf  m})
\; \nonumber
\\
- 2 \;( {\bf N} +  {\bf K} -  i{\bf A}) ({\bf L} + {\bf M} +i{\bf
B}  )   \;. \label{D.14b}
\end{eqnarray}

\noindent Besides, with the notation
\begin{eqnarray}
[kn] =  k_{0} n_{0} + {\bf n} {\bf  k} \;, \qquad  [ml]   =  m_{0}
l_{0} + {\bf l} {\bf  m}\; , \nonumber
\end{eqnarray}

\noindent relation (\ref{D.14b}) can be written as
\begin{eqnarray}
\mbox{det}\; G =
 (kk) \; (mm) + (nn)\;(ll) + 2\; [kn]\;[ml]
\nonumber
\\
- 2 \;( {\bf N} +  {\bf K} -  i{\bf A}) ({\bf L} + {\bf M} +i{\bf
B}  )   \;. \label{D.14c}
\end{eqnarray}

\noindent Remembering the designation
\begin{eqnarray}
k_{0}\; {\bf n}  = {\bf N} \; ,\qquad m_{0}\;  {\bf l} = {\bf L}
\; , \nonumber
\\
n_{0} \; {\bf k} = {\bf K} \; , \qquad l_{0} \; {\bf m} = {\bf M}
\; , \nonumber
\\
{\bf k}  \times  {\bf n}  = {\bf A } \; , \qquad {\bf m}  \times
{\bf l}  = {\bf B } \; , \label{D.15}
\end{eqnarray}

\noindent eq.  (\ref{D.14c})  take the form

\begin{eqnarray}
\mbox{det}\; G =
 (kk) \; (mm) + (nn)\;(ll) + 2\; [kn]\;[ml]\;  -
\nonumber
\\
- 2 \;( \;k_{0}\;{\bf n} +  n_{0} \; {\bf k} -  i \;{\bf k} \times
{\bf n} \;) \; (\; m_{0} \;{\bf l} + l_{0}\;{\bf m} +i\; {\bf m}
\times  {\bf l} \; )   \;. \label{D.16a}
\end{eqnarray}

\noindent In turn, eq.  (\ref{D.14a}) takes the form
\begin{eqnarray}
\mbox{det}\; G =
 (kk) \; (mm) + (nn)\;(ll) +     2 \; (  k_{0} n_{0} + {\bf n} {\bf  k}   )\; (  m_{0} l_{0} + {\bf l} {\bf  m})
\nonumber
\\
-2 \;k_{0} m_{0} \;  ({\bf n}   {\bf l})  - 2 \; k_{0} l_{0} \; (
{\bf n} {\bf m })  - 2 \;n_{0} m_{0}\; ( {\bf k} {\bf l})  -2\;
n_{0} l_{0} \; ( {\bf k} {\bf m} )
 - 2 \; ({\bf k}  \times  {\bf n} )\;
({\bf m}  \times  {\bf l}) \nonumber
\\
- 2i  k_{0}  \;{\bf n} ({\bf m}  \times  {\bf l})
  - 2i\; n_{0}  {\bf k}  ( {\bf m}  \times  {\bf l} )    +
  2im_{0}  \; ({\bf k}  \times  {\bf n} )  {\bf l} + 2i l_{0} \; ({\bf k}  \times  {\bf n} )  {\bf m}  \; .
\label{D.16}
\end{eqnarray}

\noindent Allowing for cyclic symmetry, the later can be changed
to
\begin{eqnarray}
\mbox{det}\; G =
 (kk) \; (mm) + (nn)\;(ll)
\nonumber
\\
 +  2 \; (  k_{0} n_{0} + {\bf n} {\bf  k}   )\; (  m_{0} l_{0} + {\bf l} {\bf  m})
   -2 \;k_{0} m_{0} \;  ({\bf n}   {\bf l})  -  2 \; k_{0} l_{0} \; ( {\bf n} {\bf m })
   \nonumber
\\
     -
2 \;n_{0} m_{0}\; ( {\bf k} {\bf l})  -2\; n_{0} l_{0} \; ( {\bf
k} {\bf m} ) \; -
 2 \; ({\bf k}  \times  {\bf n} )\;
({\bf m}  \times  {\bf l}) \nonumber
\\
+ 2i \; [\; +  k_{0}  \;{\bf l} ({\bf m}  \times  {\bf n})  +
m_{0}  \; {\bf k} ({\bf n}  \times  {\bf l} )
 +  l_{0} \;  {\bf k} ({\bf n}  \times  {\bf m} )    +  \; n_{0}  {\bf l}  ( {\bf m}  \times {\bf k} ) \;] \; .
\label{D.17}
\end{eqnarray}

Now, one should compare eq. (\ref{D.17}) with eq. (\ref{C.6b}):
\begin{eqnarray}
\mbox{det} \; G =
  (kk)\; (mm) + (ll) \; (nn)
\nonumber
\\
+ 2\;  (mk) \; (ln)  +2 \; (lk)\; (nm) -2\; (nk)\; (lm)  +
 4 ({\bf k} {\bf n} ) \; ({\bf m} {\bf l}) -  4 ({\bf k} {\bf m}
) \; ({\bf n} {\bf l}) \nonumber
\\
+\;  2\; i \;[\; k_{0} \;  {\bf l}  ({\bf m} \times {\bf n}) \;+\;
m_{0} \;  {\bf k}  ({\bf n} \times {\bf l}) \;+\; l_{0} \;  {\bf
k}  ({\bf n} \times {\bf m}) \;+\; n_{0} \;  {\bf l}  ({\bf m}
\times {\bf k})   \; ] \;  , \label{D.18}
\end{eqnarray}

\noindent they are the same only if
\begin{eqnarray}
  2 \; (  k_{0} n_{0} + {\bf n} {\bf  k}   )\; (  m_{0} l_{0} + {\bf l} {\bf  m})
   -2 \;k_{0} m_{0} \;  ({\bf n}   {\bf l})  - 2 \; k_{0} l_{0} \; ( {\bf n} {\bf m })
\nonumber
\\
     -
2 \;n_{0} m_{0}\; ( {\bf k} {\bf l})  -2\; n_{0} l_{0} \; ( {\bf
k} {\bf m} ) -
  2 \; ({\bf k}  \times  {\bf n} )\; ({\bf m}  \times  {\bf l})
\nonumber
\\
= 2\;  (mk) \; (ln)  +2 \; (lk)\; (nm) -2\; (nk)\; (lm) \nonumber
\\
+
 4 \; ({\bf k} {\bf n} ) \; ({\bf m} {\bf l}) -  4  \; ({\bf k} {\bf m}
) \; ({\bf n} {\bf l}) \; . \label{D.19}
\end{eqnarray}

\noindent For the right part we have
\begin{eqnarray}
\underline{\mbox{The right}} = 2 (m_{0}k_{0} - {\bf m}{\bf k})
(l_{0}n_{0} - {\bf l}{\bf n}) \nonumber
\\
+ 2 (l_{0}k_{0} - {\bf l}{\bf k})  (n_{0}m_{0} - {\bf n}{\bf m}) -
\nonumber
\\
-2 (n_{0}k_{0} - {\bf n}{\bf k})  (l_{0}m_{0} - {\bf l}{\bf m})
\nonumber
\\+
 4 ({\bf k} {\bf n} ) \; ({\bf m} {\bf l}) -  4 ({\bf k} {\bf m}
) \; ({\bf n} {\bf l}) \nonumber
\\
= 2 m_{0} k_{0} n_{0} m_{0} -2  m_{0} k_{0}  ({\bf l}{\bf n}) -
2l_{0} n_{0} ( {\bf m} {\bf k}) + 2({\bf m}{\bf k}) ({\bf l}{\bf
n}) \nonumber
\\
+ 2 l_{0}k_{0}\; n_{0}m_{0} - 2l_{0}k_{0}  ({\bf n}{\bf m}) - 2
n_{0}m_{0}  ({\bf l}{\bf k}) + 2({\bf l}{\bf k})( {\bf n}{\bf m})
\nonumber
\\
- 2 l_{0}k_{0}\; n_{0}m_{0}+ 2n_{0}k_{0} ({\bf l}{\bf m}) +
2l_{0}m_{0} ({\bf n}{\bf k}) - 2({\bf n}{\bf k}) ({\bf l}{\bf m})
\nonumber
\\
+
 4 ({\bf k} {\bf n} ) \; ({\bf m} {\bf l}) -  4 ({\bf k} {\bf m}
) \; ({\bf n} {\bf l})\; , \nonumber
\end{eqnarray}

\noindent that is
\begin{eqnarray}
\underline{\mbox{The right}}  = 2 m_{0} k_{0} n_{0} m_{0} -2
m_{0} k_{0}  ({\bf l}{\bf n}) - 2l_{0} n_{0} ( {\bf m} {\bf k}) -
2({\bf m}{\bf k}) ({\bf l}{\bf n}) \nonumber
\\
 - 2l_{0}k_{0}  ({\bf n}{\bf m}) - 2 n_{0}m_{0}  ({\bf l}{\bf k}) +
2({\bf l}{\bf k})( {\bf n}{\bf m}) \nonumber
\\
+ 2n_{0}k_{0} ({\bf l}{\bf m}) +  2l_{0}m_{0} ({\bf n}{\bf k}) +
2({\bf n}{\bf k}) ({\bf l}{\bf m}) \; . \label{D.20}
\end{eqnarray}

\noindent Taking in mind the identity
\begin{eqnarray}
 - 2 \; ({\bf k}  \times  {\bf n} )\;
({\bf m}  \times  {\bf l}) = -2 \; \epsilon_{abc} k_{b} n_{c} \;
\epsilon_{aps} m_{p} l_{s} \nonumber
\\
= -2 \; ( \delta_{bp}  \;  \delta_{cs}   - \delta _{bs}  \; \delta
_{cp} ) \; k_{b} n_{c}  \;  m_{p} l_{s} = -2   \; ({\bf k}{\bf m})
\; ({\bf n}{\bf l}) + 2\;   ({\bf k}{\bf l}) \; ({\bf n}{\bf m})
\; , \nonumber
\end{eqnarray}

\noindent for the left part
\begin{eqnarray}
\underline{\mbox{The left } } =  2 k_{0} n_{0} m_{0} l_{0} + 2
k_{0} n_{0} ({\bf l} {\bf  m}) + 2  m_{0} l_{0} ({\bf n} {\bf  k}
) + 2 ({\bf n} {\bf  k}   ) ({\bf l} {\bf  m}) \nonumber
\\
 -2 \;k_{0} m_{0} \;  ({\bf n}   {\bf l})  - 2 \; k_{0} l_{0} \; ( {\bf n} {\bf m })  -
2 \;n_{0} m_{0}\; ( {\bf k} {\bf l})  -2\; n_{0} l_{0} \; ( {\bf
k} {\bf m} ) \nonumber
\\
-2   \; ({\bf k}{\bf m}) \; ({\bf n}{\bf l}) + 2\;   ({\bf k}{\bf
l}) \; ({\bf n}{\bf m}) \; . \label{D.21}
\end{eqnarray}

\noindent Indeed, equations (D.20) and (D.21)  are the same.

However, the most simple form is (see  (\ref{D.16a}):
\begin{eqnarray}
G   =   \left | \begin{array}{rr}
k_{0}  + \; {\bf k} \; \vec{\sigma}  \;\;&  n_{0}  - \; {\bf n}  \; \vec{\sigma} \\[3mm]
- l_{0}  - \; {\bf l} \; \vec{\sigma} \;\;  & m_{0} - \; {\bf m}
\; \vec{\sigma}
\end{array} \right | \; ,
\nonumber
\\
\mbox{det}\; G =
 (kk) \; (mm) + (nn)\;(ll) + 2\; [kn]\;[ml]
 \nonumber
 \\
- 2 \;( \;k_{0}\;{\bf n} +  n_{0} \; {\bf k} -  i \;{\bf k} \times
{\bf n} \;) \; (\; m_{0} \;{\bf l} + l_{0}\;{\bf m} +i\; {\bf m}
\times  {\bf l} \; )   \; . \label{D.22}
\end{eqnarray}

Let us  consider several particular cases.

\begin{center}
{\bf Variant   A}
\end{center}
\begin{eqnarray}
+ {\bf l} \rightarrow i {\bf l} \;  , \qquad {\bf n} \rightarrow i
{\bf n}\;  , \qquad {\bf m} \rightarrow i {\bf m} \; , \qquad {\bf
k} \rightarrow i {\bf k} \;  , \nonumber
\\
G   =   \left | \begin{array}{rr}
k_{0}  + \; {\bf k} \; \vec{\sigma}  \;\;&  n_{0}  - \; {\bf n}  \; \vec{\sigma} \\[3mm]
- l_{0}  - \; {\bf l} \; \vec{\sigma} \;\;  & m_{0} - \; {\bf m}
\; \vec{\sigma}
\end{array} \right |  \rightarrow
\left | \begin{array}{rr}
k_{0}  + i\; {\bf k} \; \vec{\sigma}  \;\;&  n_{0}  - i\; {\bf n}  \; \vec{\sigma} \\[3mm]
- l_{0}  - i\; {\bf l} \; \vec{\sigma} \;\;  & m_{0} - i\; {\bf m}
\; \vec{\sigma}
\end{array} \right |\; ,
\nonumber
\\[2mm]
\mbox{det}\; G =
 [kk] \; [mm] + [nn]\;[ll] + 2\; (kn)\;(ml)
\nonumber
\\
+ 2 \;( \; k_{0}\;{\bf n} +  n_{0} \; {\bf k} +  {\bf k}  \times
{\bf n} \;) \; ( \; m_{0} \;{\bf l} +  l_{0}\;{\bf m} -  {\bf m}
\times  {\bf l} \; )   \; . \label{D.23}
\end{eqnarray}

\begin{center}
{\bf Variant   B}
\end{center}
\begin{eqnarray}
m_{a} = k^{*}_{a} \; , \qquad l_{a} = n^{*}_{a} \;  , \nonumber
\\
G (k,n)  =   \left | \begin{array}{rr}
k_{0}  + \; {\bf k} \; \vec{\sigma}  \;\;&  n_{0}  - \; {\bf n}  \; \vec{\sigma} \\[3mm]
- n^{*}_{0}  - \; {\bf n}^{*} \; \vec{\sigma} \;\;  & k^{*}_{0} -
\; {\bf k}^{*} \; \vec{\sigma}
\end{array} \right | \; ,
\nonumber
\\[2mm]
\mbox{det}\; G =
 (kk) \; (k^{*}k^{*}) + (nn)\;(n^{*}n^{*}) + 2\; [kn]\;[k^{*} n^{*}]\;  -
\nonumber
\\
- 2 \;( \;k_{0}\;{\bf n} +  n_{0} \; {\bf k} -  i \;{\bf k} \times
{\bf n} \;) \; (\; k^{*}_{0} \;{\bf n}^{*} + n^{*}_{0}\; {\bf
k}^{*}  +i\; {\bf k}^{*}  \times  {\bf n}^{*} \; )   \; .
\label{D.24b}
\end{eqnarray}

\begin{center}
{\bf Variant   C}
\end{center}
\begin{eqnarray}
m_{0} = + k_{0} \; , \qquad  l_{0} = n_{0} \;  , \qquad {\bf m} =
- {\bf k}  \; , \qquad  {\bf l} = - {\bf n} \; , \nonumber
\\
G   = \left | \begin{array}{rr}
k_{0}  + i\; {\bf k} \; \vec{\sigma}  \;\;&  n_{0}  - i\; {\bf n}  \; \vec{\sigma} \\[3mm]
- n_{0}  + i\; {\bf n} \; \vec{\sigma} \;\;  & k_{0} + i\; {\bf k}
\; \vec{\sigma}
\end{array} \right |
\nonumber
\\[2mm]
\mbox{det}\; G =
 [kk]^{2} + [nn]^{2}  + 2\; (kn)^{2} \;  -
 2 \;( \; k_{0}\;{\bf n} +  n_{0} \; {\bf k} +  {\bf k}  \times
{\bf n} \;)^{2} \;  . \label{D.26b}
\end{eqnarray}

\end{document}